\documentclass[preprint]{aastex63}

\usepackage{color}

\revised{\today}
\submitjournal{ApJ}

\shorttitle{Inverse Compton Cooling}
\shortauthors{Kinch et al.}

\begin{document}

\title{Inverse Compton Cooling in the Coronae of Simulated Black Hole Accretion Flows}

\correspondingauthor{Brooks E. Kinch}
\email{kinch@lanl.gov}

\author[0000-0002-8676-425X]{Brooks E. Kinch}
\affiliation{CCS-2: Computational Physics and Methods \\
Los Alamos National Laboratory \\
NM 87545, USA}

\author[0000-0003-3547-8306]{Scott C. Noble}
\affiliation{Gravitational Astrophysics Laboratory \\
NASA Goddard Space Flight Center \\
Greenbelt, MD 20771, USA}

\author[0000-0002-2942-8399]{Jeremy D. Schnittman}
\affiliation{Gravitational Astrophysics Laboratory \\
NASA Goddard Space Flight Center \\
Greenbelt, MD 20771, USA}

\author[0000-0002-2942-8399]{Julian H. Krolik}
\affiliation{Department of Physics and Astronomy \\
Johns Hopkins University \\
Baltimore, MD 21218, USA}

\begin{abstract}

We present a formulation for a local cooling function to be employed in the diffuse, hot corona region of 3D GRMHD simulations of accreting black holes. This new cooling function calculates the cooling rate due to inverse Compton scattering by considering the relevant microphysics in each cell in the corona and approximating the radiation energy density and Compton temperature there by integrating over the thermal seed photon flux from the disk surface. The method either assumes ion and electron temperatures are equal (1T), or calculates them separately (2T) using an instantaneous equilibrium approach predicated on the actual relevant rate equations (Coulomb and Compton). The method is shown to be consistent with a more detailed ray-tracing calculation where the bulk of the cooling occurs, but is substantially less costly to perform. As an example, we apply these methods to a \textsc{harm3d} simulation of a $10 M_\odot$, non-spinning black hole, accreting at nominally 1\% the Eddington value. Both 1T and 2T approaches lead to increased radiative efficiency and a larger fraction of total cooling in the corona as compared to the original target-temperature cooling function used by \textsc{harm3d}, especially in the 1T case. Time-averaged post-processing reveals that the continuum spectral observations predicted from these simulations are qualitatively similar to actual X-ray binary data, especially so for the 1T approach which yields a harder power-law component ($\Gamma = 2.25$) compared to the 2T version ($\Gamma = 2.53$).

\end{abstract}

\keywords{Magnetohydrodynamical simulations (1966) --- General relativity (641) --- Accretion (14) --- X-ray binary stars (1811)}

\section{Introduction}

Astrophysical black hole accretion is a profoundly difficult problem to simulate realistically. The rich and varied observational signatures from both supermassive and stellar-mass black holes arise from the interplay of General Relativity (GR), magnetohydrodynamics (MHD), and radiative processes. A proper treatment must be at least finely enough resolved in space and time to capture the evolution of the magneto-rotational instability \citep{bal91a, haw91a}---the core underlying accretion mechanism---on the appropriate background metric (generally, Kerr). The first code to achieve this in a global (albeit 2D) context was \textsc{harm} \citep{gam03a}, which has formed the basis of many subsequent codes \citep{eht19a} including \textsc{harm3d} \citep{nob09a}, the global 3D GRMHD code we extend and apply in this work. The final component of the physics, radiation, has proved the most difficult to fully incorporate into any code.

Simultaneously solving the MHD equations \emph{and} the global angle- and energy-dependent radiation transport equation, in General Relativity, is both computationally expensive (typically prohibitively so) and technically challenging. Even so, significant progress has been made in the last decade, though the problem is usually made tractable by introducing at least one of the following simplifying assumptions: abandoning General Relativity in favor of a pseudo-Newtownian description of the gravitational potential, while performing realistic, multi-angle group radiation transport \citep{jia14a, jia14b, jia19a, jia19b}; limiting the possible angular-dependence of the radiation field by invoking either flux-limited diffusion \citep{zan11a, roe12a} or, more recently, the ``M1 closure'' relation, in either axisymmetric (2D) \citep{sad14a} or 3D simulations \citep{fra12a, fra14a, mck14a, sad16a}; or Monte Carlo \citep{rya15a}/hybrid MC techniques \citet{rya19a}. Most attempts have eschewed energy-dependent transfer in favor of a ``grey'' atmosphere---the radiation field is treated as monochromatic, coupled to the fluid only through the Rosseland mean opacity \citep{ryb86a}. The first of these approximations, the pseudo-Newtonian potential, is especially problematic in regions close to the black hole where General Relativistic effects play a critical role in determining both the structure of the accretion flow and photon trajectories. The others are essentially variants of a diffusion approximation, and are best suited to the cooler, denser, and optically thick body of the accretion disk, where the environment is similar to those found in stellar atmospheres---the field from which these methods, and grey transfer, originate \citep{cha60a}. With the exception of Monte Carlo methods \citep{rya15a, rya19a}, these are all especially poorly-suited to the diffuse, hot, optically thin corona, especially at small radii near the black hole.

\textsc{harm3d} has employed a local cooling function approach to emulate the effects of radiative cooling in moderately accreting (0.01--0.3 Eddington) systems: gravitationally-bound gas hotter than a specified target temperature is cooled to the target temperature over one orbital timescale; the target temperature is chosen so as to achieve a desired small disk aspect ratio, i.e., a geometrically thin disk. In this paper, we employ a cooling function which instead considers the relevant microphysics within each fluid element and from there calculates the rate at which the internal energy of the gas is converted to photons. This calculation requires, in general, knowledge of the energy-dependent radiation field in all cells. Our new cooling function is specifically tailored to the diffuse corona regime---it is designed to improve the realism of the coronal domain without introducing the additional \emph{substantial} overhead of full transport. We achieve this by employing a series of simplifying but enabling assumptions, which we check against a Monte Carlo radiation transport calculation. The new cooling function is ``switched on'' at a time after the simulation has evolved long enough with the original, target-temperature cooling function that it has achieved a statistically steady-state, and applies only in the corona. Within the disk body, the original, target-temperature cooling function remains in place. In the second part of this paper, we also explore the consequences of weak ion-electron coupling in the corona, in the context of our more physically-motivated cooling function. Throughout, we examine the effects of the new cooling function on the dynamical, thermodynamic, and spectral properties of a fiducial $10 M_\odot$, non-spinning ($a = 0$) black hole simulation accreting at approximately 1\% Eddington.

\section{Inverse Compton Cooling Function}

The target-temperature cooling function was introduced in \citet{nob09a} and further developed in \citet{nob10a} and \citet{nob11a}. A radius-dependent target temperature $T_*$ (expressed in \textsc{harm3d}'s dimensionless code units) is set according to

\begin{equation}
T_* = \frac{\pi}{2} \frac{R_z(r)}{r} \left[ \frac{H(r)}{r} \right]^2,
\end{equation}

\noindent where $R_z$ is the relativistic correction to the vertical component of gravity at radius $r$ and $H$ is the density-weighted scale height of the disk [equation from \citet{nob09a}, corrected from \citet{abr97a}]. The desired disk aspect ratio, $H/r$, is chosen \emph{a priori} to achieve a geometrically thin disk---for the starting point ThinHR simulation series used here, the desired disk aspect ratio is $H/r = 0.05$. When gas on a bound orbit exceeds the target temperature, it is cooled back to the target temperature by introducing a nonzero sink term on the right-hand side of the local stress-energy conservation equation solved by \textsc{harm3d}:

\begin{equation}
\nabla_\nu T^\nu_{\ \mu} = - \mathcal{L} u_\mu,
\label{eq:stress_energy}
\end{equation}

\noindent here $T^\nu_{\ \mu}$ is the stress-energy tensor and $u_\mu$ is the specific four-momentum; $\mathcal{L}$ is chosen so that the gas cools to the target temperature over one circular orbital period at its radius. If the gas is at or below $T_*$, or is gravitationally unbound, $\mathcal{L} = 0$. As the gas is cooled, its pressure and therefore support against gravity decreases, settling back toward the midplane and thereby achieving a geometrically thin disk.

The target-temperature approach has several key benefits: as intended, it gives rise to a geometrically thin, optically thick, relatively cool, dense disk, sandwiched between a hotter, diffuse corona---a configuration with considerable observational support \citep{haa91a}; the implementation is independent of the central black hole mass scale $M$ and the nominal accretion rate $\dot{m}$ (in Eddington units) and therefore, in principle, the results of a single simulation can be scaled to both stellar-mass X-ray binary systems and supermassive active galactic nuclei; and it is easy to evaluate as it depends only on local properties of a given fluid element.

On the other hand, of course, it is unphysical: the choice of target temperature is motivated not by the relevant microphysics, but by the desire to achieve a configuration-by-design that agrees well with observational evidence. There are other concerns as well. By virtue of its implementation as a local sink term, it is everywhere ``optically thin''; that is, the dissipation rate is an entirely \emph{intensive} quantity---even deep within the disk, energy lost (nominally to photons) simply vanishes, while in reality these photons would diffuse through the optically thick material, scattering and undergoing absorption/re-emission along the way. In the corona, gravitationally unbound matter does not cool at all, while gas that is cooled does so on a circular orbital timescale which may not relate to its actual cooling time.

An optically thin cooling function is, however, a good approximation in the truly optically thin corona. Thus we seek a more physical cooling function there while still retaining the implementation of Equation \ref{eq:stress_energy}. In addition, we understand the actual physical mechanism behind coronal cooling: the inverse Compton (IC) scattering of thermal seed photons from the disk surface off of very hot electrons---this is exactly the physics treated with great care by \textsc{pandurata} \citep{sch13b}. Below we detail the development of a new cooling function $\mathcal{L}$ to replace the target-temperature cooling function in the corona; to emphasize the physical origin of the new cooling function---and to distinguish it from the target-temperature cooling function which will remain in use in the disk body---we refer to it simply as the IC cooling function.

At first, we will require the ion and electron populations to be at the same temperature in each simulation cell, $T_e = T_i$, i.e., a one temperature fluid. In section \ref{2T_section}, we extend the method to treat the ion and electron temperatures separately, $T_e \neq T_i$, i.e., a two temperature fluid, with the assumption that electrons are heated only through Coulomb collisions with ions. These two prescriptions represent, essentially, the limiting cases of maximally- and minimally-efficient radiative cooling, respectively. The one temperature fluid assumption requires no special description of the ion-electron coupling mechanism---we simply assume that some strong coupling mechanism exists, or that turbulent dissipation is shared nearly equally between ions and electrons, or a combination thereof. Later, we will posit a specific coupling mechanism, namely Coulomb collisions, in addition to the assumption that all turbulent energy is initially injected into the ions only.

\subsection{Inverse Compton Power}

To distinguish the corona volume from the disk body, two disk photosphere surfaces, $\Theta_\mathrm{top} (r, \phi)$ and $\Theta_\mathrm{bot} (r, \phi)$, are defined by integrating the electron scattering opacity from the $z$-axis toward the midplane:

\begin{equation}
\int_0^{\Theta_\mathrm{top} (r, \phi)} \kappa \rho (r, \phi) d \theta \sqrt{g_{\theta\theta}} = 1,
\label{eq:Theta_top}
\end{equation}

\begin{equation}
-\int_\pi^{\Theta_\mathrm{bot} (r, \phi)} \kappa \rho (r, \phi) d \theta \sqrt{g_{\theta\theta}} = 1,
\label{eq:Theta_bot}
\end{equation}

\noindent where $\kappa$ is the electron scattering opacity, $0.4\ \mathrm{cm}^2\ \mathrm{g}^{-1}$. Note that if no solution exists for the above equations for a given $(r, \phi)$ such that $\Theta_\mathrm{top} < \Theta_\mathrm{bot}$, then simply no disk body exists there. For a given point $(r, \theta, \phi)$ such that $\Theta_\mathrm{top}$ and $\Theta_\mathrm{bot}$ exist, the point is considered in the disk body if $\Theta_\mathrm{top} < \theta < \Theta_\mathrm{bot}$; otherwise, the point is considered in the corona. The location of the photosphere surfaces depends on the choice of accretion rate through the overall scale of the density; as discussed in section \ref{units_and_scaling} below, a larger nominal $\dot{M}/\dot{M}_\mathrm{Edd}$ implies a larger overall $\rho$ which, all else equal, results in photospheres which lie further from the midplane. The cooling function we derive in this section is applied only to those simulation cells whose centers lie in the corona.

The classic expression for the energy exchanged per inverse Compton scatter per unit time is [see \citet{blu70a} for details on Compton scattering expressions used below]:

\begin{equation}
P_\mathrm{IC} = \frac{4}{3} \sigma_T c \gamma^2 \beta^2 u_\mathrm{rad},
\end{equation}

\noindent where $\sigma_T$ is the Thomson scattering cross section, $c$ is the speed of light, $\gamma = 1/\sqrt{1 - \beta^2}$ with $\beta = v/c$---where $v$ refers to the electron velocity---and $u_\mathrm{rad}$ is the radiation energy density. In the nonrelativistic limit, i.e., when the dimensionless electron temperature $\Theta_e \equiv k_B T_e/m_e c^2 \ll 1$, $\langle \gamma^2 \beta^2 \rangle \simeq 3 \Theta_e$ (with the averaging indicated by angle brackets performed over a thermal electron velocity distribution), and so

\begin{equation}
P_\mathrm{IC, non-relativistic} = \frac{4}{3} \sigma_T c (3 \Theta_e) u_\mathrm{rad}.
\end{equation}

\noindent In the relativistic limit, i.e., $\Theta_e \gg 1$, $\langle \gamma^2 \beta^2 \rangle \simeq 12 \Theta_e^2$, therefore

\begin{equation}
P_\mathrm{IC, relativistic} = \frac{4}{3} \sigma_T c (12 \Theta_e^2) u_\mathrm{rad}.
\end{equation}

Because each expression is much larger than the other in their appropriate regimes, we can represent the IC power in either limit by their sum. In addition, we multiply by the electron density $n_e = \chi (\rho/m_i)$, where $\chi$ is the free electron fraction (number of free electrons per ion, equal to 1.21 for a fully-ionized plasma with solar elemental abundances; a variable $\chi$ might also be used to account for the presence of electron-positron pairs due to pair production) and $m_i$ is the mean ion mass ($\simeq m_p$), to translate from energy exchanged per scatter per unit time to a volumetric cooling rate. The final expression is:

\begin{equation}
\mathcal{L}_\mathrm{IC} = \frac{4 \sigma_T c \chi}{m_i} \rho u_\mathrm{rad} \Theta_e (1 + 4 \Theta_e).
\label{eq:L_IC}
\end{equation}

This is the expression for the IC cooling rate which enters into \textsc{harm3d}'s stress-energy equation \ref{eq:stress_energy} in place of the target-temperature cooling rate for those fluid elements in the corona. It requires as input: the density, the electron temperature, and the radiation energy density. As discussed, we will assume for now that some strong coupling mechanism forces $T_e = T_i$. From standard thermodynamics and the ideal gas law, the gas pressure $p_\mathrm{gas}$, the internal energy density $u$, and the electron and ion temperatures are related by

\begin{equation}
p_\mathrm{gas} = (c_P/c_V - 1) u = n_e k_B T_e + n_i k_B T_i,
\label{eq:ideal_gas_law}
\end{equation}

\noindent where $c_P/c_V$ is the ratio of specific heats (the adiabatic index) [equal to 5/3 for a monatomic gas; though we assume $c_P/c_V = 5/3$ for simplicity in this work, a more detailed consideration of relativistic plasma physics leads to a slightly different, and variable, adiabatic index \citep{mig07a}]. From this we derive an expression for the dimensionless electron temperature:

\begin{equation}
\Theta_e = \frac{m_i}{m_e} \frac{c_P/c_V - 1}{1 + \chi} \frac{u}{\rho c^2}.
\label{eq:Theta_e}
\end{equation}

\noindent This expression depends only on the (mass) density and the internal energy density---which, like the density, is part of \textsc{harm3d}'s fluid solution. Again, this equation holds so long as there is some strong coupling mechanism forcing $T_e = T_i$.

In order to estimate the radiation energy density $u_\mathrm{rad}$ at each point in the corona, we first make several key simplifications:

\begin{enumerate}

\item We ignore general and special relativistic effects. The thermal seed photons launched from the disk surface are assumed to travel in straight rays, undergoing neither red/blue-shifting nor beaming due to the bulk fluid flow of the rotating accretion disk, nor gravitational redshifting due to their origin in a deep gravitational potential.

\item We do not account for scattering or obscuration of the disk emission by intervening corona material between the disk surface and a given point in the corona.

\item We adopt the ``fast light'' approximation, i.e., we do not account for the light travel time between a point on the disk photosphere and a point in the corona; rather, the radiation energy density in the corona each time step is computed from the thermal flux from the disk surface at the same time step.

\end{enumerate}

With these assumptions in place, we derive an expression for $u_\mathrm{rad}$ by integrating the thermal seed photon flux over the disk surface with an appropriate geometric weight. Let $\mathbf{r}$ indicate the location of the coronal cell in question, and let $\mathbf{r}^\prime$ locate a surface cell on the photosphere. Then:

\begin{equation}
d u_\mathrm{rad} (\mathbf{r}) = \frac{1}{c} \frac{F_\mathrm{disk} (\mathbf{r}^\prime) \cos \vartheta d A^\prime}{R^2},
\label{eq:du_rad}
\end{equation}

\noindent where the factor of $1/c$ translates flux into energy density, $\mathbf{R} = \mathbf{r} - \mathbf{r}^\prime$, $\vartheta$ is the angle between $\mathbf{R}$ and the disk surface normal vector $\hat{\mathbf{n}}$, $dA^\prime$ is the (infinitesimal) element of the disk surface area, and $F_\mathrm{disk}$ is the assumed blackbody seed photon flux with effective temperature $T_\mathrm{eff}$, set by integrating the (target-temperature) cooling function within the disk at the specified $(r^\prime, \phi^\prime)$:

\begin{equation}
\int_{\Theta_{\mathrm{top}}}^{\Theta_{\mathrm{bot}}} \mathcal{L} d\theta \sqrt{g_{\theta\theta}} = 2 \sigma_{\mathrm{SB}} T_{\mathrm{eff}}^4.
\label{eq:T_eff}
\end{equation}

\noindent We express $\cos \vartheta$ in terms of the spherical coordinates of the coronal and photosphere cells:

\begin{equation}
\cos \vartheta = \pm \frac{r}{R} \left[ \sin \theta \cos \theta^\prime \left( \cos \phi \cos \phi^\prime + \sin \phi \sin \phi^\prime \right) - \cos \theta \sin \theta^\prime \right] = \frac{r}{R} G_\pm ( \mathbf{r}, \mathbf{r}^\prime ),
\label{eq:costheta}
\end{equation}

\noindent where $+$ is used for the lower half of the corona and $-$ for the upper half. In addition, it can be shown that the infinitesimal solid angle $d\Omega^\prime$ subtended by the disk surface element at $\mathbf{r}^\prime$ from the coronal cell at $\mathbf{r}$ is:

\begin{equation}
d\Omega^\prime = 2 \pi \left[ 1 - \left( 1 + \frac{dA^\prime}{\pi R^2} \right)^{-1/2} \right].
\label{eq:dOmega}
\end{equation}

\noindent Substituting equations \ref{eq:costheta} and \ref{eq:dOmega} into equation \ref{eq:du_rad}, and integrating over the disk surface, we arrive at

\begin{equation}
u_\mathrm{rad} (\mathbf{r}) = 2\pi \int_{\partial A^\prime} \frac{r}{R} F_\mathrm{disk} (\mathbf{r}^\prime) G_\pm (\mathbf{r}, \mathbf{r}^\prime) \left[ 1 - \left( 1 + \frac{dA^\prime}{\pi R^2} \right)^{-1/2} \right].
\label{eq:u_rad}
\end{equation}

\noindent This expression is computed in each coronal cell. To ease the computational burden of the numerical integration of the right-hand side of the above equation, a coarsened sampling of the photosphere grid (e.g., including only every eighth $\phi$ grid-point and every sixth $r$ grid-point) is used without significant loss in accuracy. Furthermore, $u_\mathrm{rad}$ does not need to be computed every time step: \textsc{harm3d} fluid update time steps are guaranteed to be sufficiently short compared to the thermal time scale that the integral in equation \ref{eq:u_rad} varies very little from one time step to the next. In practice, we have verified that evaluating $u_\mathrm{rad}$ only once every 20 time steps introduces $< 1\%$ error.

\subsection{The Compton Temperature}

It is also useful to estimate the Compton temperature, $T_C$, in each coronal cell. The Compton temperature---the temperature at which Compton heating is balanced by Compton cooling---is, in the non-relativistic limit:

\begin{equation}
k_B T_C = \frac{1}{4} \frac{\int_0^\infty h \nu J_\nu d\nu}{\int_0^\infty J_\nu d\nu} = \frac{1}{4} \langle \varepsilon \rangle,
\label{eq:T_C}
\end{equation}

\noindent where $J_\nu$ is the mean intensity at frequency $\nu$. In other words, the Compton temperature is equal to one quarter the mean photon energy (so defined). For a pure blackbody, it is easy to show that $\langle \varepsilon \rangle = 3.832 kT_\mathrm{eff}$. Therefore, $T_C$ in a given coronal cell is found by averaging $T_\mathrm{eff}$ over the disk surface, weighted by the contribution of each particular photosphere surface element's flux to the total radiation energy density. That is:

\begin{equation}
T_C = \frac{3.832}{4} \frac{\sum_n u_{\mathrm{rad},n} T_{\mathrm{eff},n}}{\sum_n u_{\mathrm{rad},n}},
\label{eq:T_C_v2}
\end{equation}

\noindent where $T_{\mathrm{eff},n}$ is the effective temperature of the $n^\mathrm{th}$ disk surface element and $u_{\mathrm{rad},n}$ is the evaluation of the right-hand side of equation \ref{eq:u_rad} for a particular photosphere element (performed in the course of numerical integration). The expression for IC power derived above, equation \ref{eq:L_IC}, is valid only if $T_e \gg T_C$; otherwise, an additional term for Compton heating is required:

\begin{equation}
\mathrm{Compton\ heating} = \frac{\sigma_T}{m_e c} n_e u_\mathrm{rad} \langle \varepsilon \rangle.
\label{eq:comp_heat}
\end{equation}

As we show below, the $T_e \gg T_C$ condition is always met using the 1T assumption (and is still fairly well satisfied under 2T), which indicates that Compton heating is negligible compared to Compton cooling.

\subsection{The IC Cooling Time}

Equation \ref{eq:L_IC} for the IC cooling power is the time-rate change in the internal energy of the gas. That is, in the fluid rest frame:

\begin{equation}
\frac{du}{dt} = -\mathcal{L}_\mathrm{IC}.
\end{equation}

\noindent Substitute the expression for $\Theta_e$ in equation \ref{eq:Theta_e} (derived assuming $T_e = T_i$) into equation \ref{eq:L_IC} to solve the differential equation above:

\begin{equation}
u(t) = \frac{u_0}{(1 + b u_0) e^{at} - b u_0},
\label{eq:u_of_t}
\end{equation}

\noindent in which

\begin{eqnarray}
a \equiv \frac{4 \sigma_T}{m_e c} \frac{\chi}{1 + \chi} \left(c_P/c_V - 1\right) u_\mathrm{rad}, \\\relax
b \equiv 4 \frac{m_i}{m_e} \frac{\left(c_P/c_V - 1\right)}{1 + \chi} \frac{1}{\rho c^2}.
\end{eqnarray}

\noindent From equation \ref{eq:u_of_t}, we calculate the cooling time $t_\mathrm{cool}$, or the time over which---assuming $u_\mathrm{rad}$ and $\rho$ are constant---the internal energy decreases from $u_0 \to u_0/e$:

\begin{equation}
t_\mathrm{cool} = \frac{1}{a} \ln \left( \frac{e + b u_0}{1 + b u_0} \right).
\end{equation}

By inspection of the above equations, we see that the conditions for a short cooling time are either a high radiation energy density or a high initial electron temperature. Because of the quadratic term in the expression for the IC cooling rate, using the instantaneous rate of equation \ref{eq:L_IC} in \textsc{harm3d} can overestimate the cooling in very hot cells. We instead define the cooling over one time step by appropriately time-averaging the cooling rate during the time step:

\begin{equation}
\bar{\mathcal{L}_\mathrm{IC}} = \frac{u_0 - u(\Delta \tau)}{\Delta \tau},
\end{equation}

\noindent where $\Delta \tau$ is the proper time interval in the given coronal cell corresponding to the \emph{global} simulation (coordinate) time step $\Delta t$. In addition, if $t_\mathrm{cool} < \Delta \tau$ in any cell, the global time step is reset accordingly to match the shortest $t_\mathrm{cool}$. As mentioned above, the value of $\bar{\mathcal{L}_\mathrm{IC}}$ is set assuming $u_\mathrm{rad}$ and $\rho$ are constant. While $u_\mathrm{rad}$ is a function of (an average over) the disk structure---and will therefore vary more slowly---a given coronal cell's density can of course change rapidly. To maintain the integrity of the numerical fluid dynamics solution, we must be sure that no cells cool too substantially each time step.

In practice, $\bar{\mathcal{L}_\mathrm{IC}}$ differs from $\mathcal{L}_\mathrm{IC}$ as given in equation \ref{eq:L_IC}, and $t_\mathrm{cool} < \Delta \tau$, only briefly right after the new corona cooling function is first ``switched on.'' Because the target-temperature cooling function only cools bound gas (and does so less efficiently, as we see below), the corona cools rapidly under the new regime; the usual \textsc{harm3d} time step determination procedure is generally sufficient after several-to-ten $M$ of simulation time have elapsed.

\subsection{A Note on Units and Scaling}
\label{units_and_scaling}

The derivations in the previous sections used physical, cgs units. To implement these equations in \textsc{harm3d}, however, we must translate to code units. Using notation such that a quantity $x$ is converted from code to cgs units by $x_\mathrm{cgs} = [x] x_\mathrm{code}$, we rewrite equation \ref{eq:L_IC} as:

\begin{equation}
[\mathcal{L}] \mathcal{L}_\mathrm{IC,code} = \frac{4 \sigma_T c \chi}{m_i} [\rho] \rho_\mathrm{code} [u_\mathrm{rad}] u_\mathrm{rad,code} \Theta_e (1 + 4 \Theta_e).
\label{eq:L_IC_v2}
\end{equation}

\noindent Note that, in units for which $G = c = 1$, $\Theta_e$ (already a dimensionless quantity) is trivially re-expressed in code units by setting $c = 1$, $u \to u_\mathrm{code}$ and $\rho \to \rho_\mathrm{code}$ in equation \ref{eq:Theta_e}. Consulting the conversion factors for $\mathcal{L}$ and $\rho$ from \citet{sch13a}, we have:

\begin{eqnarray}
[\mathcal{L}] = \frac{4 \pi c^7}{\kappa G^2 M^2} \frac{\dot{m} / \eta}{\dot{M}_{\mathrm{code}}}, \label{eq:L_2}\\\relax
[\rho] = \frac{4 \pi c^2}{\kappa G M} \frac{\dot{m} / \eta}{\dot{M}_{\mathrm{code}}}, \label{eq:rho_2}\\\relax
[u_\mathrm{rad}] = [u] = c^2 [\rho].
\end{eqnarray}

\noindent Substituting these expressions into equation \ref{eq:L_IC_v2}, we find:

\begin{equation}
\mathcal{L}_\mathrm{code} = \left( \frac{16 \pi \sigma_T \chi}{m_i \kappa} \right) \frac{\dot{m} / \eta}{\dot{M}_\mathrm{code}} \rho_\mathrm{code} u_\mathrm{rad,code} \Theta_e (1 + 4 \Theta_e).
\label{L_code}
\end{equation}

\noindent The term in parentheses is dimensionless and, assuming a fixed free electron fraction (totally-ionized plasma is essentially guaranteed in the corona), constant.

The other dimensionless term, $\left(\dot{m}/\eta\right)/\dot{M}_\mathrm{code}$, does \emph{not} appear in the ``code units'' form of the expression for the target-temperature cooling rate, but it does appear in the code units form of the IC cooling rate expression; it appears also in the expressions \ref{eq:L_2}, \ref{eq:rho_2} for translating the cooling rate and density from code to cgs units. This term serves to set the overall mass (and therefore dissipation) scale of the accretion flow. The radiative efficiency $\eta$ is defined such that, in cgs units, the bolometric luminosity is related to the mass accretion rate by $L = \eta \dot{M} c^2$. The value of $\eta$ used in the above expressions, however, must be chosen \emph{a priori}. For ease of comparison with analytic accretion disk theory, we choose the \citet{nov73a} values for the nominal radiative efficiency; for $a = 0$, $\eta_\mathrm{NT} = 0.0572$. The NT radiative efficiencies are calculated assuming the fluid elements of an axisymmetric, time-steady disk radiate the entirety of their gravitational binding energy as measured from the ISCO radius. By definition, an accretion flow characterized by some radiative efficiency, say, $\eta_\mathrm{NT}$, accreting at $\dot{m} = 0.01$, has a luminosity equal to 0.01 $L_\mathrm{Edd}$. We describe these choices as ``nominal'' because their purpose is not to specify a resulting luminosity, but simply to set a scale---because the radiative efficiency may not correspond exactly to the ISCO binding energy, but is computed by the simulation, the actual luminosity should be on order of, but is not expected to be exactly equal to, 0.01 $L_\mathrm{Edd}$.

Whereas the scale-free nature of the target-temperature cooling function required choosing in advance only the dimensionless spin $a/M$, this is no longer the case when using the IC corona cooling function. The nominal accretion rate $\dot{m}$ appears in the expression for $\mathcal{L}_\mathrm{code}$, as described above; also, the very first step when using this new cooling function is to divide the disk and corona by calculating the location of the photospheres---because $\rho \propto \dot{m}$, the upper and lower photosphere surfaces move further from the midplane with increasing $\dot{m}$, decreasing the volume of the simulation space governed by the IC corona cooling function. The structure of the accretion geometry therefore depends on the choice of $\dot{m}$.

For a chosen $\dot{m}$, however, the simulation results are still scalable with $M$. To see why this remains possible, consider how each term in equation \ref{eq:L_IC} scales with $M$: $\rho \propto M^{-1}$ and $u_\mathrm{rad} \propto F_\mathrm{disk} \propto \mathcal{L} r \propto M^{-2} M$; therefore $\mathcal{L}_\mathrm{IC} \propto M^{-2}$. Thus the IC corona cooling function has the same scaling with $M$ as does the target-temperature cooling function. The Compton temperature scales with $M$ like so: $T_C \propto T_\mathrm{eff,disk} \propto (\mathcal{L} r)^{1/4} \propto M^{-1/4}$. The scaling is weak, and such that $T_C$ \emph{decreases} for more massive black holes. The condition for the validity of the IC cooling rate, equation \ref{eq:L_IC}, is $T_e = (m_e c^2/k_B) \Theta_e \gg T_C$. As discussed earlier, $\Theta_e$ is a dimensionless quantity that does not scale with $M$ or $\dot{m}$ (it is proportional to the ratio of two quantities with identical scaling relationships, $\rho$ and $u$). Therefore, because $T_e \gg T_C$ is satisfied for stellar-mass black holes (as we show below), it is necessarily satisifed for supermassive black holes.

\subsection{Uncooled Material}
\label{uncooled_material}


\textsc{harm3d}'s inversion routines are susceptible to a close subtraction error when recovering the gas pressure from the total pressure in regions which are magnetically-dominated [because $p_B \approx p_\mathrm{total}$, the ``positive pressure problem'' \citep{bal99a}]. Numerical errors can result in large, artificial pressure gradients across adjacent cells, rapidly accelerating material in gross violation of energy conservation. As a remedy, \textsc{harm3d} instead solves an \emph{entropy} conservation equation where solution of the stress-energy conservation equation (\ref{eq:stress_energy}) fails; that is:

\begin{equation}
\nabla_\mu \left( \mathcal{S} u^\mu \right) = 0,
\label{eq:entropy}
\end{equation}

\noindent where $\mathcal{S} \equiv p/\rho^{c_P/c_V - 1}$. Cells subject to evolution using the entropy equation do not obey the energy conservation equation, and are therefore unaffected by the specified cooling function; the effect is small \citep{nob09a}, however, and the region of the simulation volume to which it is applied ultimately contributes little to the overall simulation dynamics or X-ray observables. The entropy conservation equation is employed if \emph{either} $B^2/\rho$ (the magnetization) or $B^2/u$ (approximately the reciprocal of the plasma $\beta$) exceed certain critical values. For the simulations we show below, the thresholds chosen are $B^2/u > 10^4$, $B^2/\rho > 1$.

\section{Application of the Inverse Compton Cooling Function}

To demonstrate the IC cooling function with strongly coupled ions and electrons ($T_e = T_i$), we apply it to the zero spin ThinHR simulation, after the system has already evolved for $10,000 M$ with the original, target-temperature cooling function applied everywhere. For our example case, the (scale free) ThinHR simulation is scaled to a $10 M_\odot$ central black hole with a nominal accretion rate of 1\% Eddington ($\dot{m} = 0.01$). The simulation is continued for $2000M$ in two versions: with the IC cooling function in use in the corona, and with the original, target-temperature cooling function remaining in use everywhere. We refer to the time at which at the IC cooling function switches on as $t = 0$ (even when referencing the target-temperature everywhere simulation).

\begin{figure}
\epsscale{0.85}
\plotone{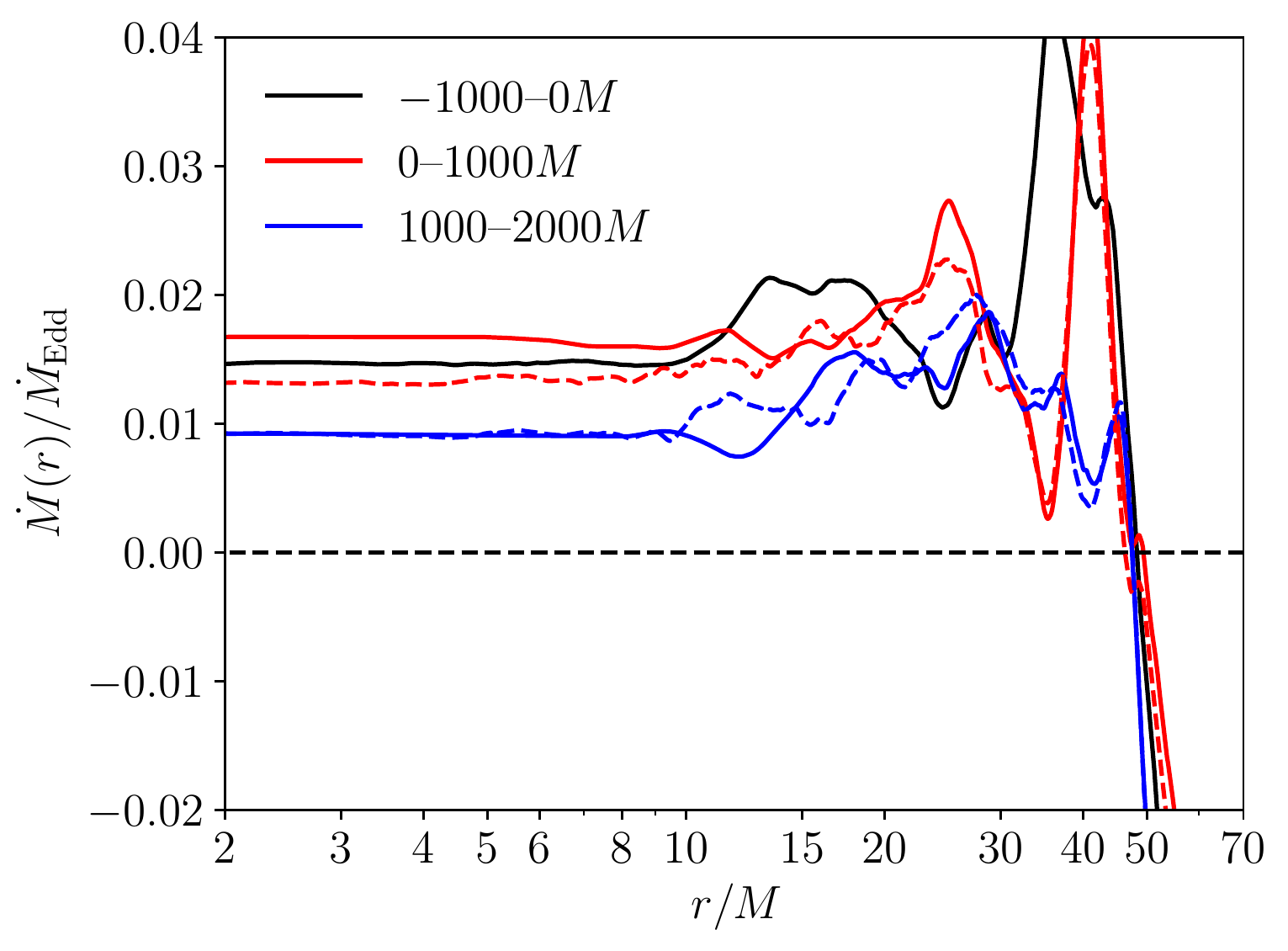}
\caption{The shell-integrated mass inflow rate, $\dot{M}$, as a function of radial coordinate, averaged in time over three $1000 M$ windows, expressed in ratio to the nominal Eddington mass accretion rate. The dashed curves represent the continuation of the target-temperature everywhere simulation, while the solid red and blue curves are for the run where the IC cooling function is switched on at $t = 0$. \label{fig:inflow}}
\end{figure}

One of the chief assumptions of analytic accretion disk theory \citep{sha73a} is that the flow is time-steady; this implies mass inflow equilibrium, i.e., the rate of inward mass flow through shells at all radii is the same and does not vary with time. In a real or simulated system the accretion flow is turbulent and therefore highly variable in time and space; however, inflow equilibrium can still be defined in a time-averaged sense. The shell-integrated mass inflow rate as a function of the radial coordinate is \citep{nob12a}

\begin{equation}
\dot{M}(r) = -4 \int_0^\pi d\theta \int_0^{\pi/2} d\phi \ \rho u^r \sqrt{-g};
\label{eq:Mdot}
\end{equation}

\noindent the factor of 4 and the azimuthal integration bounds are necessary as these simulations are performed over only one quadrant. In Figure \ref{fig:inflow}, we show the radial-dependence of the mass inflow rate---expressed in ratio to the nominal Eddington mass accretion rate---time-averaged over three windows: the last $1000 M$ of the ThinHR simulation, and the first and second $1000 M$ intervals after. For comparison, the same data is shown for the continuation past $t = 0$ for the original, target-temperature everywhere version.

From Figure \ref{fig:inflow} it is apparent that these simulations achieve time-averaged inflow equilibrium out to $r \sim 15 M$; though the \emph{value} of the mass accretion rate (most sensibly measured at the event horizon at $r = 2 M$) does vary. The ThinHR simulations are initialized with a finite amount of matter available to accrete (they are not ``fed'' by a companion as a real X-ray binary would be), some of which must be pushed out to larger radii as other material sheds its angular momentum and moves inward. In practice, it is challenging to design a simulation that achieves total radius- and time-independent inflow equilibrium while remaining computationally feasible. Nevertheless, the region interior to $r \sim 15 M$ accounts for about half of the total cooling in the simulation volume, and is the origin of the most important observable X-ray diagnostics.

\begin{figure}
\epsscale{0.85}
\plotone{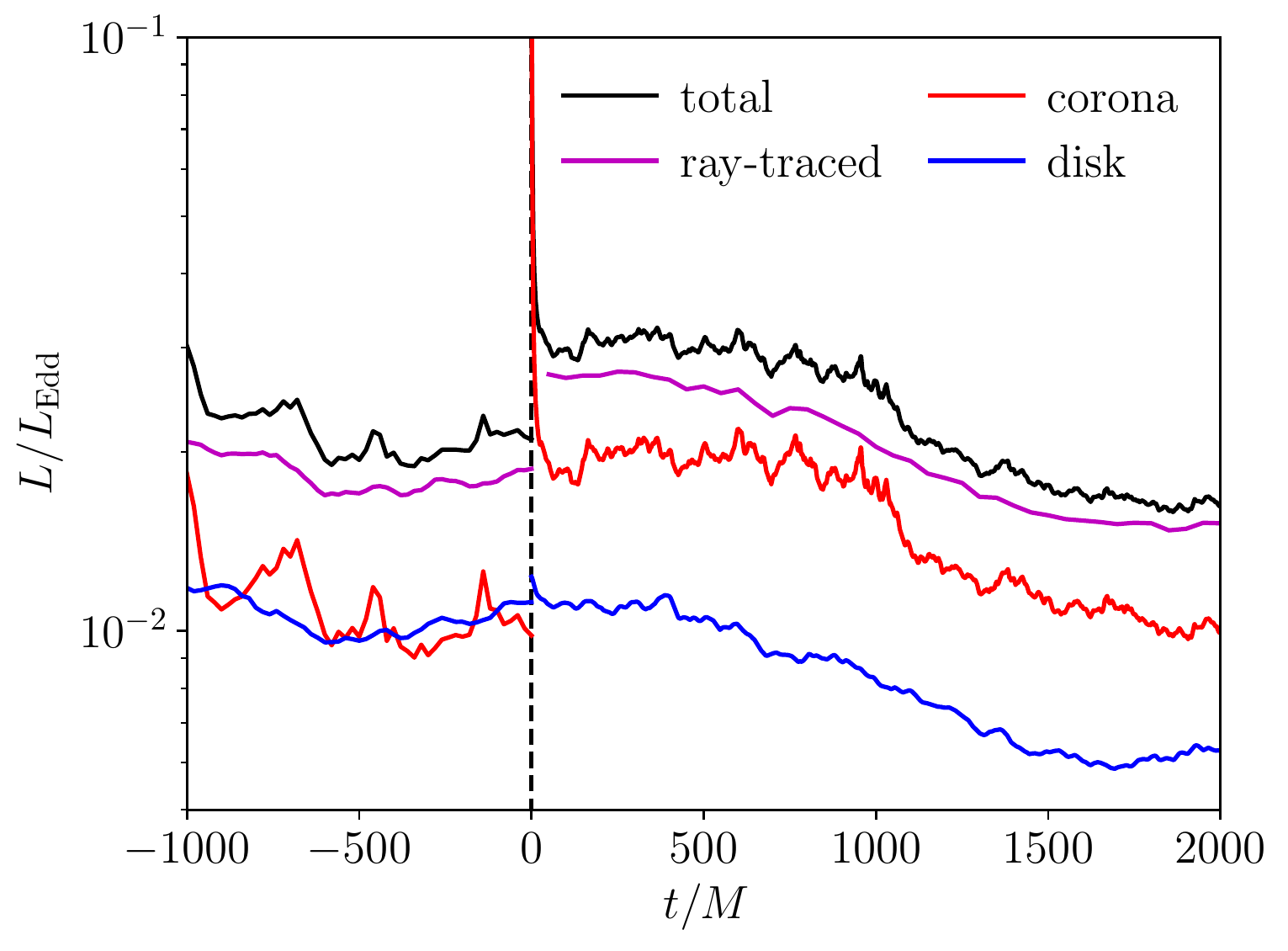}
\caption{The total cooling rate, the contributions from the disk and corona separately, and the luminosity measured by the distant observer via the ray-tracing code \textsc{pandurata}, as fractions of the Eddington luminosity, as functions of time. The IC cooling function is switched on (in the corona only) at $t = 0$. \label{fig:cooling}}
\end{figure}

The volume-integrated total cooling rate is

\begin{equation}
L_\mathrm{tot} = 4 \int_{R_h}^{R_\mathrm{max}} dr \int_0^\pi d\theta \int_0^{\pi/2} d\phi \ \mathcal{L} u^t \sqrt{-g},
\label{eq:L_int}
\end{equation}

\noindent where $\mathcal{L}$ is the fluid rest frame value of the IC cooling function (in the corona) or the target-temperature cooling function (in the disk). $R_h = 2 M$ is the radius of the event horizon, and $R_\mathrm{max} = 70 M$ is the outer radial boundary of the simulation volume. Figure \ref{fig:cooling} shows the total cooling rate, as well as the contribution from the disk and corona separately, expressed in ratio to the Eddington luminosity, as functions of time---including the last $1000 M$ of the ``starter'' simulation, with the division between disk and corona superimposed. With the old cooling function, the disk accounted for nearly exactly half of the total cooling; with the new cooling function, the mean disk fraction is 0.38 and nearly constant in time---even though the overall luminosity decreases over the length of the simulation.

Note that $\mathcal{L}$ is defined in the fluid rest frame, therefore $L$ as defined does not account for all special and General Relativistic effects, nor for the capture of photons by the black hole. We therefore use the Monte Carlo ray-tracing code \textsc{pandurata} \citep{sch13b, sch13a}, applied to successive snapshots of the \textsc{harm3d} simulation, to calculate the (post-processed) bolometric luminosity that reaches infinity. As is apparent from the figure, equation \ref{eq:L_int} provides a consistent overestimation of the more careful ray-tracing calculation.

As discussed in section \ref{units_and_scaling}, the conversion from code units to physical cgs units requires specification of the radiative efficiency $\eta$. We choose the analytic accretion disk theory \citep{nov73a} values for convenience. However, we can also compute the simulation's radiative efficiency post hoc by comparing the time-averaged accretion rate (as measured at the event horizon) to the time-averaged luminosity as measured by an observer at infinity. For the $0$--$1000M$ window, this yields $\eta = 0.0929$; for the $1000$--$2000M$ window, $\eta = 0.0983$. These values are greater than both the NT value for an $a = 0$ black hole, $\eta_\mathrm{NT} = 0.0572$, and the value for the last $1000 M$ of the input simulation, $\eta = 0.0712$.

Both $L$ and $\dot{M}$ scale from code to cgs units proportional to the factor $\left(\dot{m}/\eta_\mathrm{NT}\right)/\dot{M}_\mathrm{code}$, and so the inferred radiative efficiency is nearly independent of the nominal choice. It is not \emph{entirely} independent, however, as the location of the photosphere surfaces which divide the corona from the disk depends on the physical density scale (equations \ref{eq:Theta_top} and \ref{eq:Theta_bot}). As Figure \ref{fig:cooling} indicates, the increase in inferred radiative efficiency is due to the increased magnitude of the IC cooling function compared to the target-temperature version. Measured values of the radiative efficiency are generally not observationally-accessible, though long time-averages for $L/L_\mathrm{Edd}$ are. Regardless of our choice for the overall scaling factor $\left(\dot{m}/\eta\right)/\dot{M}_\mathrm{code}$, our simulations can be just as well characterized (and thereby compared to observable systems) by $L/L_\mathrm{Edd}$. When the IC cooling function is turned on, the coronal luminosity triples, doubling $L/L_\mathrm{Edd}$ for this system. Over time, the total luminosity trends downward, returning to its nominal value but with a greater share due to the corona. While not shown in Figure \ref{fig:cooling}, the long term trend of the ``starter'' target-temperature everywhere simulation continued for $t > 0$ mirrors the IC simulation evolution. The primary difference is that the IC simulation's corona represents a consistently larger fraction of the total luminosity.

\begin{figure}
\plottwo{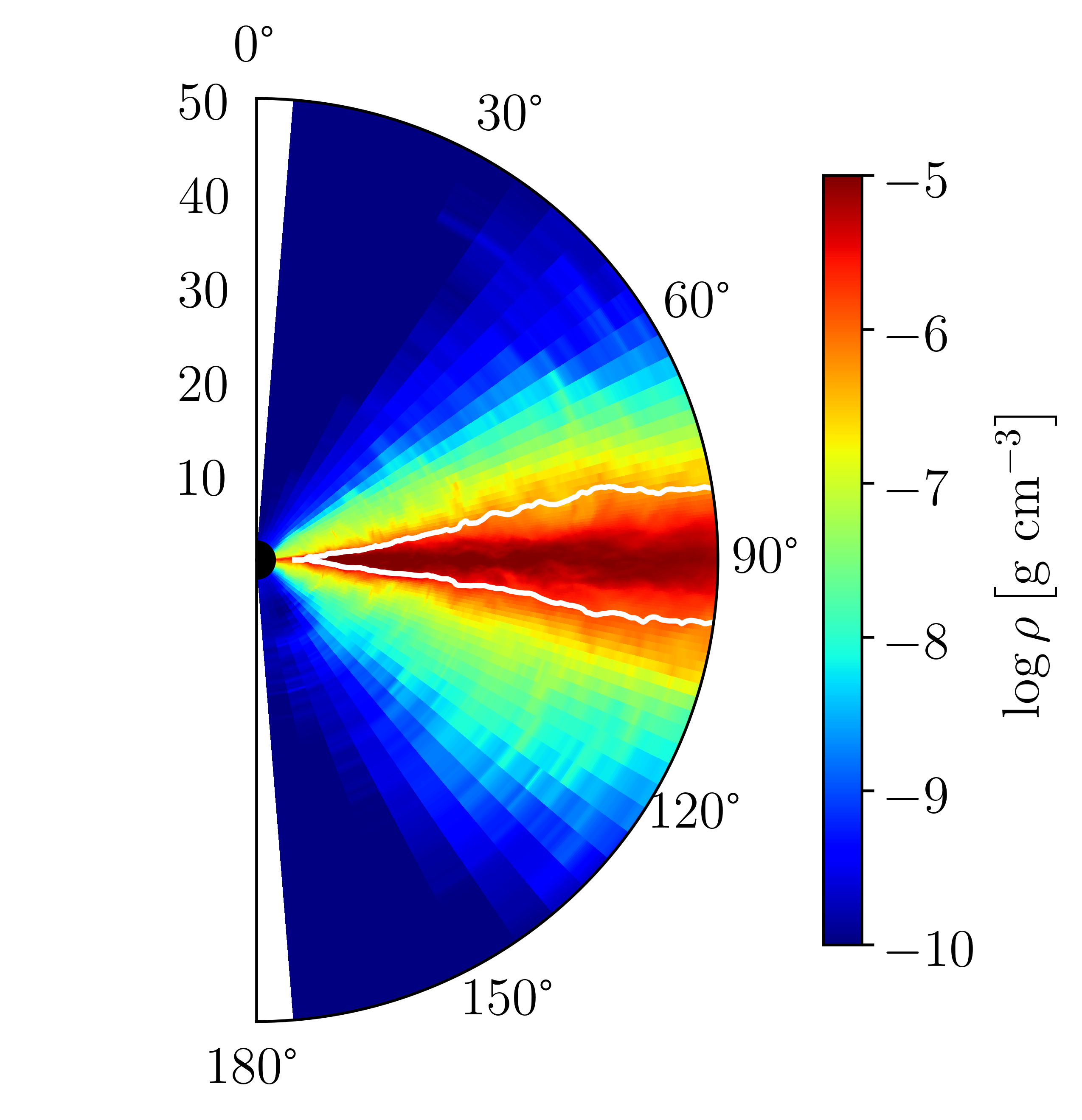}{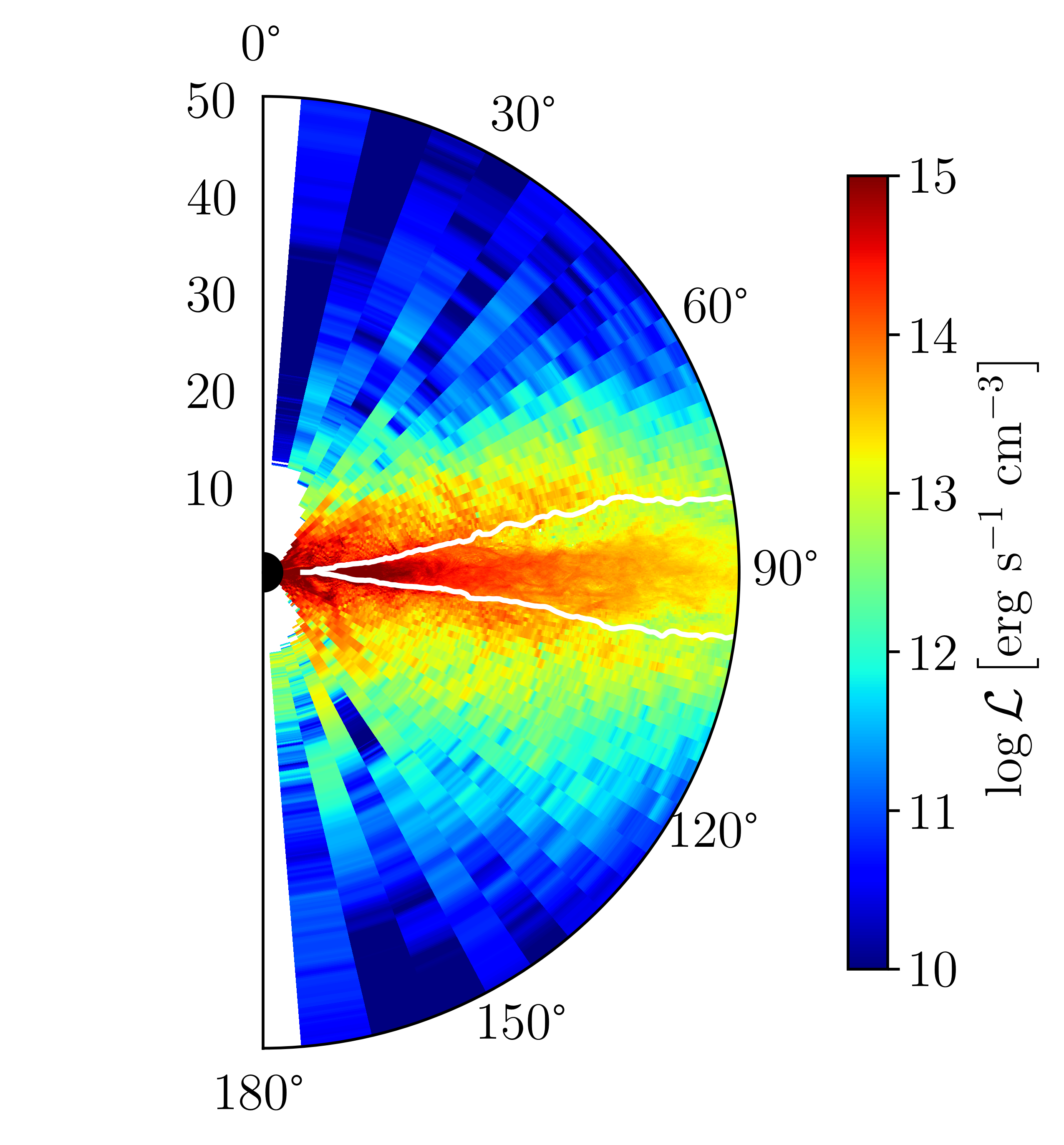}
\caption{Azimuthally-averaged values of the density (left) and the fluid frame cooling rate (right), for the snapshot at $t = 1000 M$. The white lines indicate the ($\phi$-averaged) photosphere surfaces. Cells with zero cooling are shown in white. \label{fig:rho_and_cool}}
\end{figure}

\begin{figure}
\plottwo{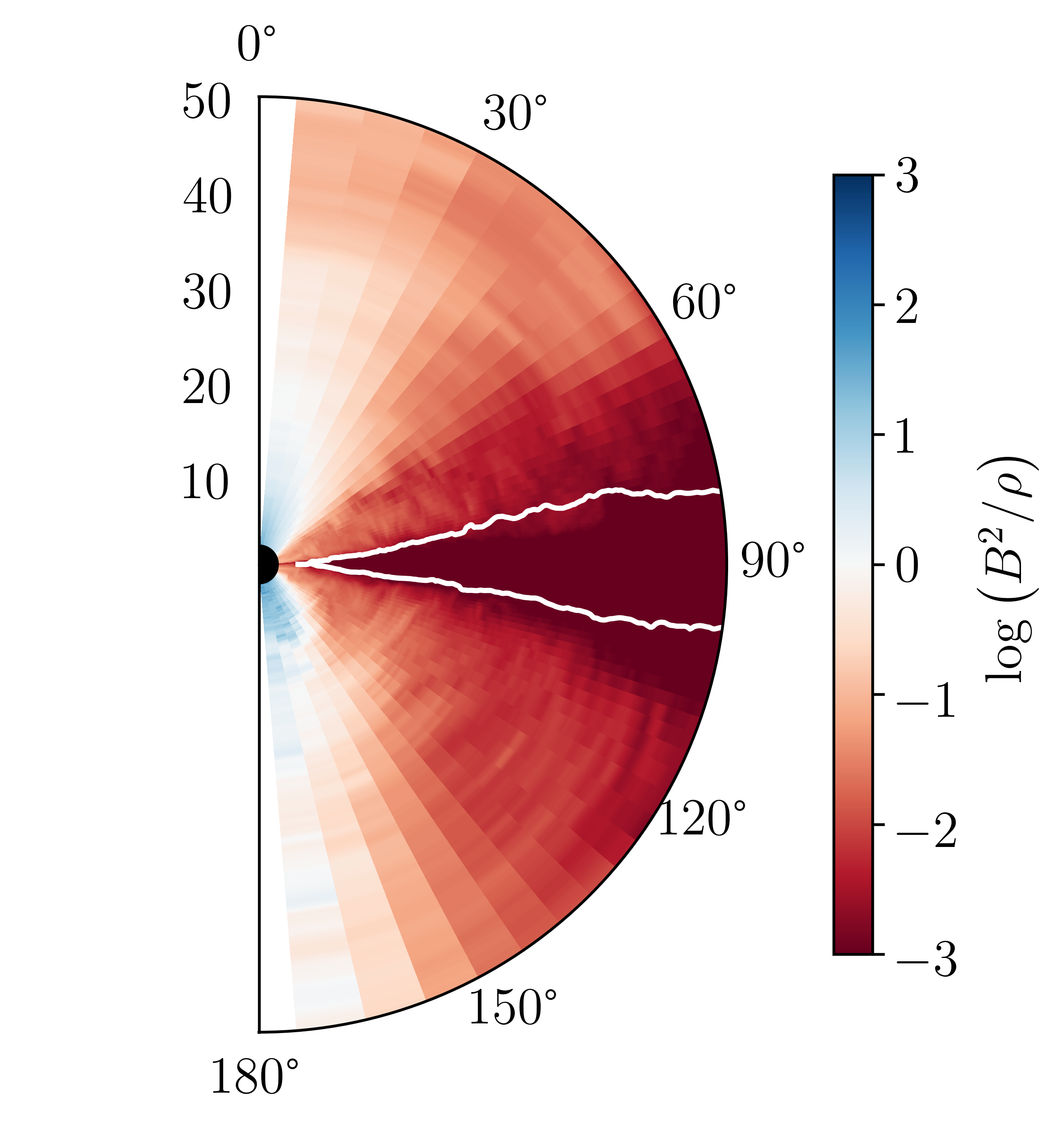}{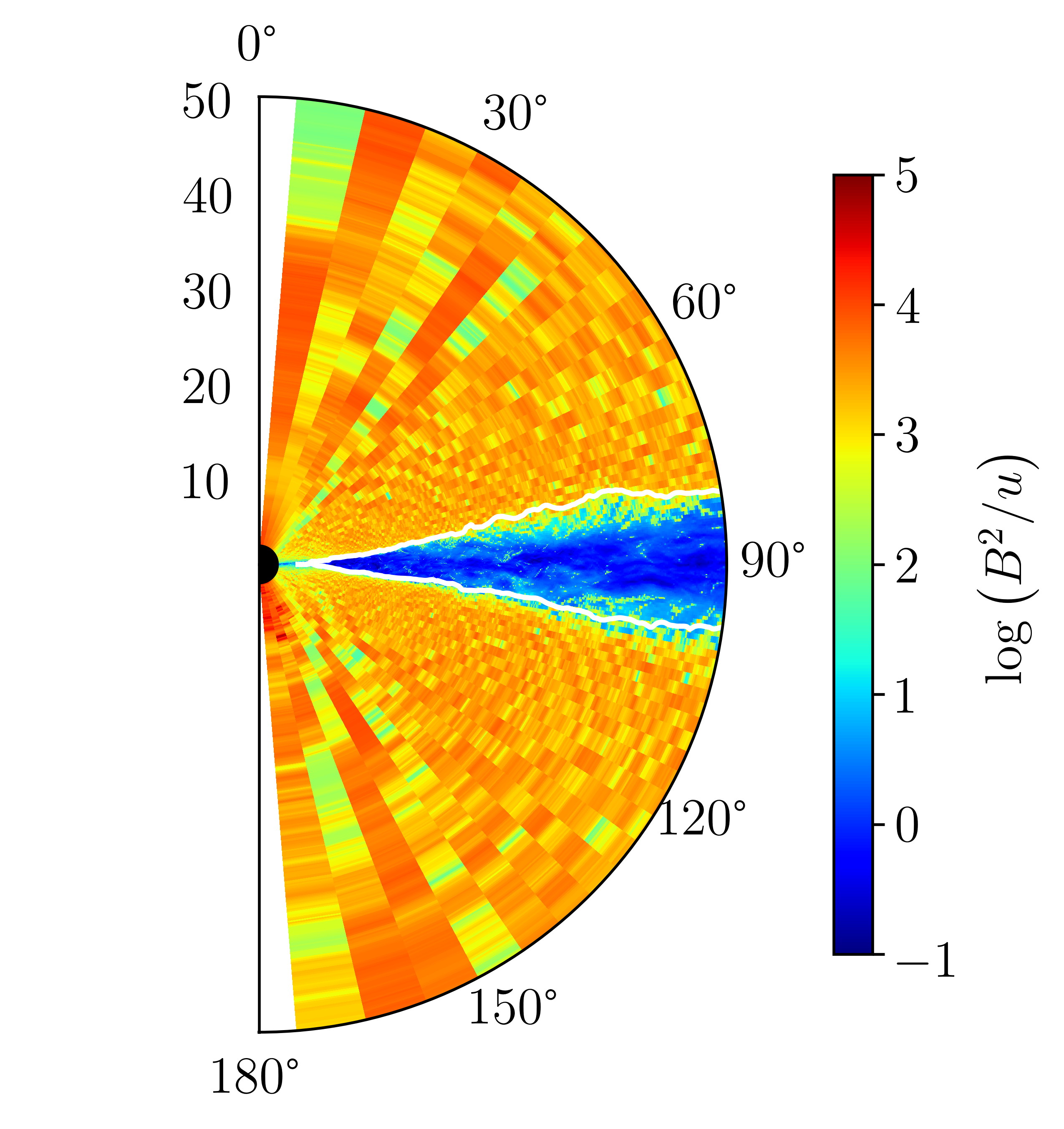}
\caption{Azimuthally-averaged values of the ratio of magnetic energy density to mass-energy density (left) and magnetic energy density to internal energy density (right), for the snapshot at $t = 1000 M$. The white lines indicate the ($\phi$-averaged) photosphere surfaces. \label{fig:bsqs}}
\end{figure}

In the figures below, we show azimuthally-averaged cross sections of various simulation quantities, all at $t = 1000 M$ for the IC simulation. Figure \ref{fig:rho_and_cool} shows the density and cooling rate for this snapshot: note the rapid decrease in density away from the midplane (at $90^\circ$). Azimuthal averages of the dimensionless ratios $B^2/\rho$ and $B^2/u$ are shown in Figure \ref{fig:bsqs}. The critical values which trigger evolution via the entropy equation (see section \ref{uncooled_material}) are met only in the relatively small region near the $z$-axis at $r \lesssim 10 M$.

\begin{figure}
\epsscale{0.85}
\plotone{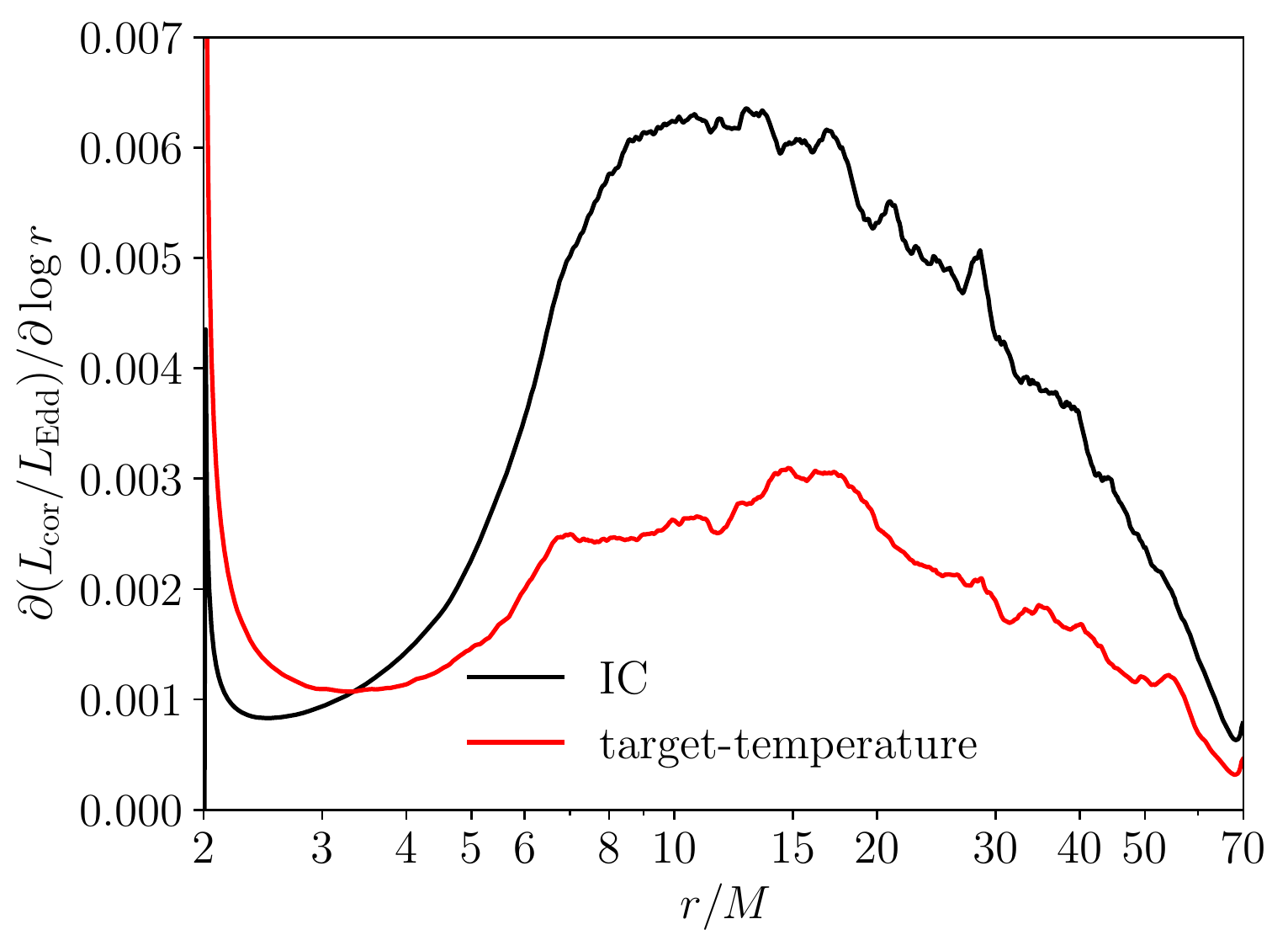}
\caption{The radial distribution of the coronal cooling power, time-averaged, for the $0$--$2000M$ interval for both the IC-cooled corona simulation and the continued target-temperature everywhere run.\label{fig:dLdr_comp}}
\end{figure}

The IC simulation's corona is nearly everywhere more luminous than is the target-temperature version's. Compare the radial distributions of coronal cooling as shown in Figure \ref{fig:dLdr_comp} (shell-integrated and time-averaged). Except for a deficit at very small radii, the IC-cooled gas radiates more energy. The consequence is that the internal energy of the gas decreases after the IC cooling function is switched on, resulting in an overall lower temperature corona (see Figure \ref{fig:dMdT_comp} below). Even though the corona is generally cooler, it radiates more energy---this is not contradictory, because the cooling rate as determined with the target-temperature method is tied to the orbital time scale, not the relevant thermodynamic time scale. The rate of advection of thermal energy through the black hole event horizon reduces by one-third after the IC cooling function is switched on, consistent with a greater fraction of injected heat being radiated away. As shown in Figure \ref{fig:inflow}, the accretion rate \emph{increases} after the IC cooling function is switched on, due to the rapidly cooling gas losing some of its pressure support against gravity; eventually, however, the IC simulation's accretion rate returns to the same value of the continued target-temperature everywhere run (though its heat advection rate remains lower).

Even after the corona has cooled off, the IC cooling function simply radiates more energy per unit time than does the target-temperature cooling function, even when supplied with the same inputs. The magnitude of corona cooling shown in Figure \ref{fig:rho_and_cool} is 15--50 times greater than the target-temperature cooling function would be for the same simulation data (the ratio is larger closer to the disk); the cooler corona at $t = 1000M$ is nearer to the target temperature, enhancing this difference.

\begin{figure}
\epsscale{0.6}
\plotone{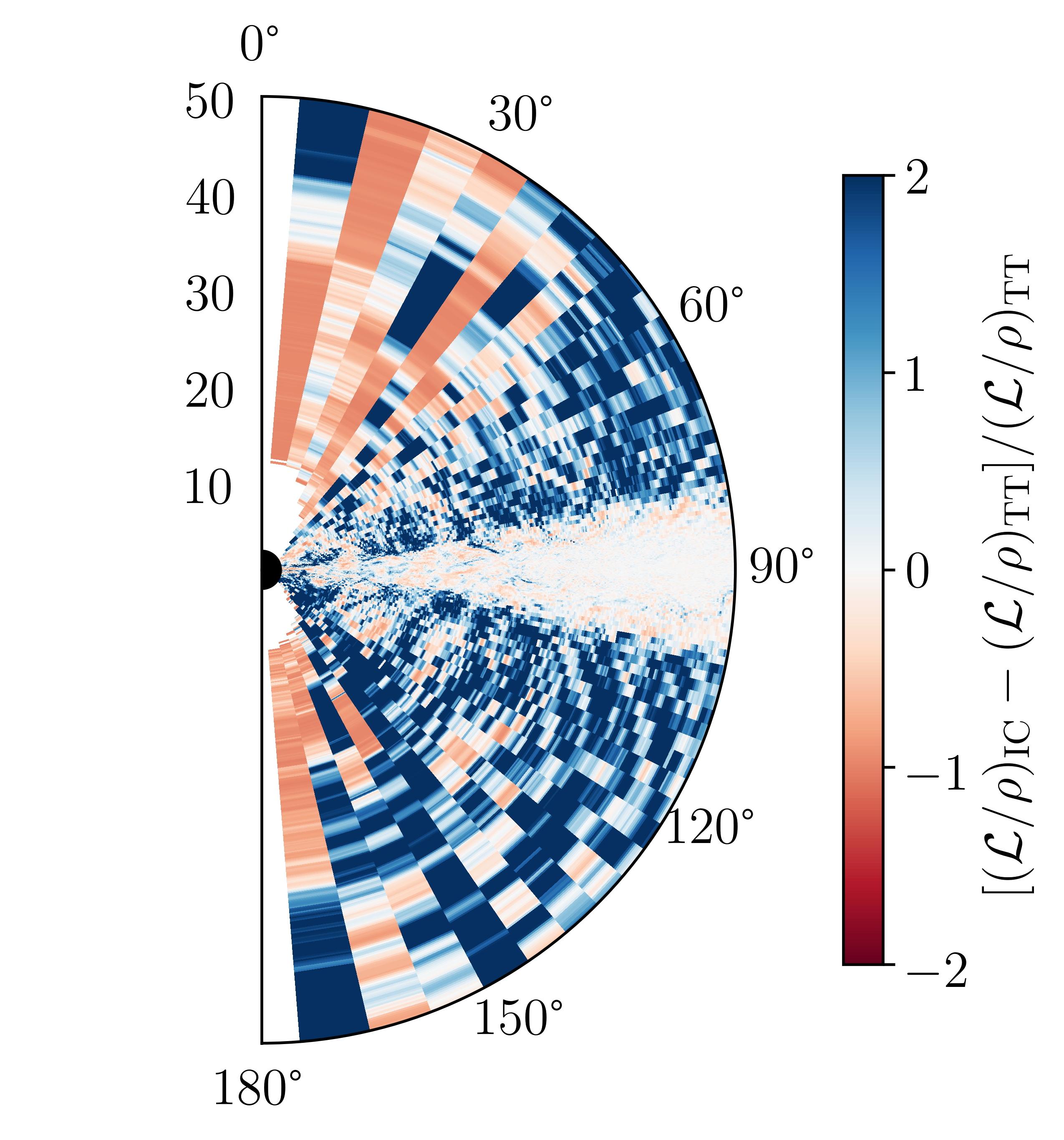}
\caption{The fractional difference between the ratio of local cooling rate to density, at $t = 1000M$, between the IC simulation and the target-temperature everywhere simualtion; azimuthally-averaged. \label{fig:cool_o_rho}}
\end{figure}

Figure \ref{fig:cool_o_rho} shows the fractional difference between the cooling rate per unit mass of the IC simulation data and the continued target-temperature everywhere run, both evaluated at $t = 1000M$. While the disk body---which remains subject to the target-temperature cooling function in both instances---is unsurprisingly nearly the same, the IC-cooled corona is, overall, more luminous.

\begin{figure}
\epsscale{0.85}
\plotone{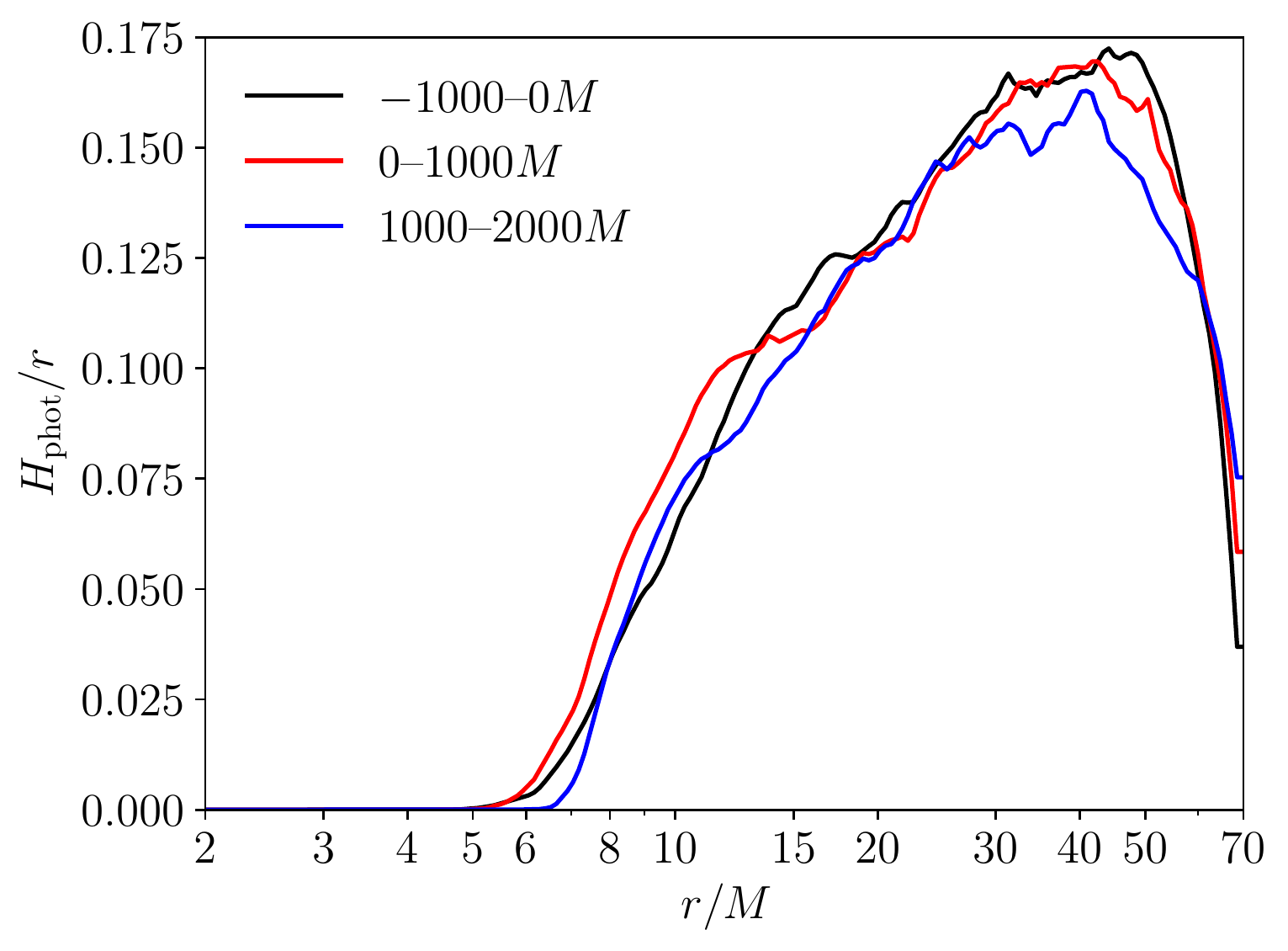}
\caption{The height of the photosphere, as a function of radius, averaged over azimuth, top and bottom, and for three time windows. While the location of the inner cutoff of the photosphere (the ``reflection edge'') varies somewhat with time, it turns out to be close to the ISCO at $r = 6M$. \label{fig:photo_height}}
\end{figure}

The corona is even more magnetically supported ($B^2/u \gg 1$ in the corona in Figure \ref{fig:bsqs}) by $t = 1000M$ than it was at the start, as the plasma $\beta$ decreases by a factor of ten near the poles and by a factor of one hundred near the disk. The magnetic field strength varies but generally increases, especially in and near the disk, and the gas pressure falls everywhere. The increase in the magnetic field strength is a trend observed both with and without switching on the IC cooling function; the large decrease in plasma $\beta$, however, is due to the falling gas pressure as the corona radiates away its internal energy more rapidly with the IC cooling function compared to the target-temperature version. The overall geometry of the accretion flow is not significantly affected, however---as indicated in Figure \ref{fig:photo_height}, the demarcation between corona and disk remains relatively fixed.

The dynamical evolution of the corona is clearly affected by the choice of cooling function. By the end of their $2000M$ runs, the IC-cooled simulation is advecting thermal energy through the event horizon at a rate of 0.001 $L_\mathrm{Edd}$; by contrast, the continued target-temperature everywhere run is advecting at a rate 0.0015 $L_\mathrm{Edd}$. At the same time, the IC simulation's volume-integrated luminosity is 0.003 $L_\mathrm{Edd}$ higher than the target-temperature run's. The decreased advective loss, therefore, accounts for only one-sixth of the increase in luminosity of the IC simulation relative to the continued target-temperature simulation. The magnetic heating rate of the coronal plasma increases as a consequence of the more realistic coronal cooling function.

\pagebreak

\section{Validation Against a Ray-Tracing Solution}

\textsc{pandurata} launches thermal seed photon packets from the disk surface and follows their trajectories through the corona until they either scatter off electrons, re-impinge on the disk surface, are captured by the black hole, or escape to an observer at infinity. When a photon packet scatters, its energy is convolved (in the fluid rest frame) with a thermal Compton scattering kernel corresponding to the presumed $T_e$ at the point of scattering [this procedure is described in \citet{kin19a}]. By following large numbers of photon packets in this manner, the net difference between each pre- and post-scatter photon packet energy is used to construct the effective fluid rest frame inverse Compton cooling rate for a given spatial map of $T_e$. Through an iterative procedure, $T_e$ is adjusted everywhere until it matches \textsc{harm3d}'s cooling map. At the same time, \textsc{pandurata} computes the energy- and inclination-dependent flux as would be measured by an observer at infinity. Because \textsc{pandurata}'s procedure and \textsc{harm3d}'s IC cooling function calculation share several key assumptions---``fast light'' and a local, thermal disk seed photon spectrum (determined for both via equation \ref{eq:T_eff})---we can isolate effects due to the ways in which they differ: \textsc{pandurata} accounts for all special and General Relativistic effects, and the occlusion of disk radiation by intervening material, while \textsc{harm3d} does not.

\begin{figure}
\plottwo{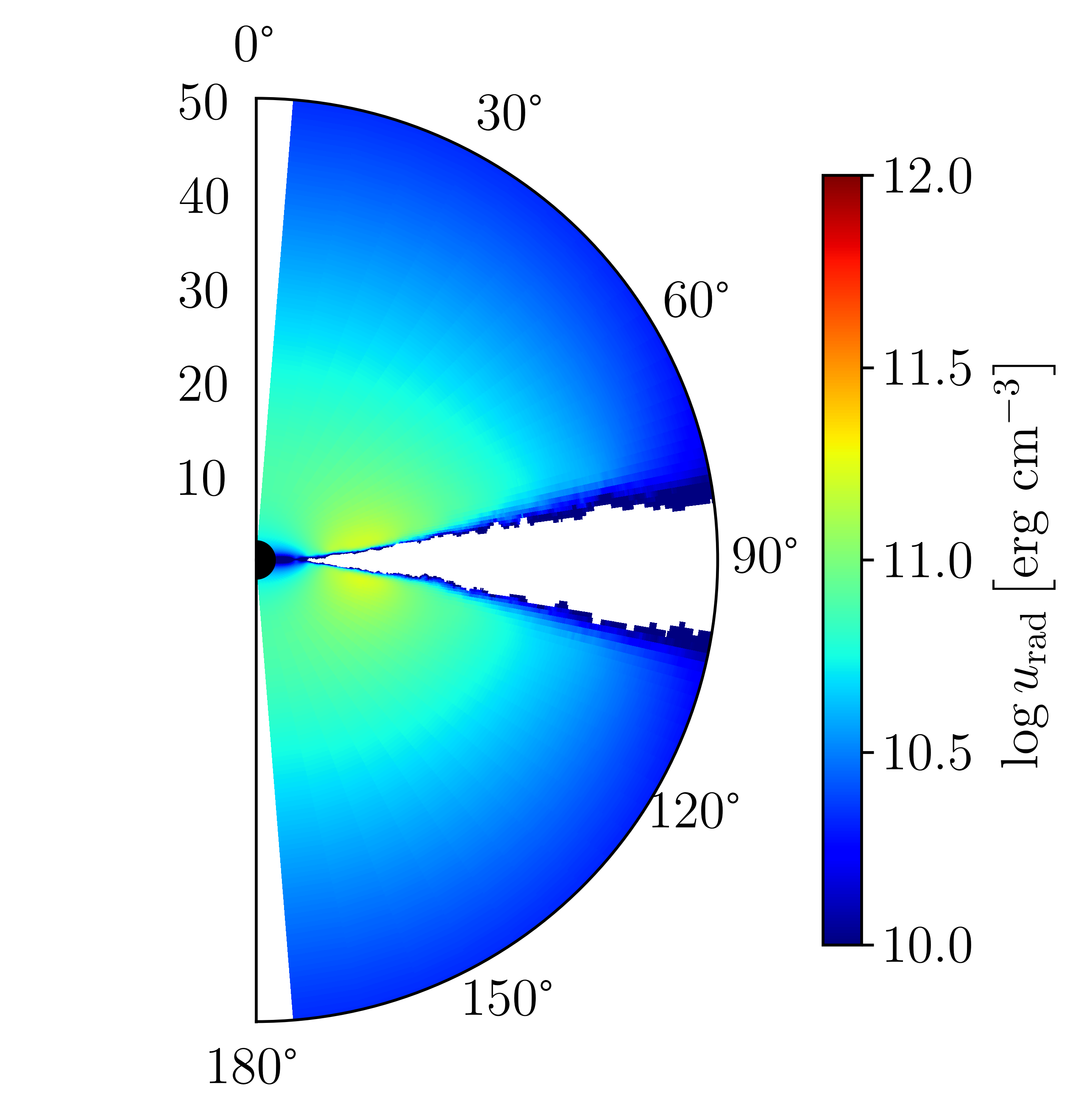}{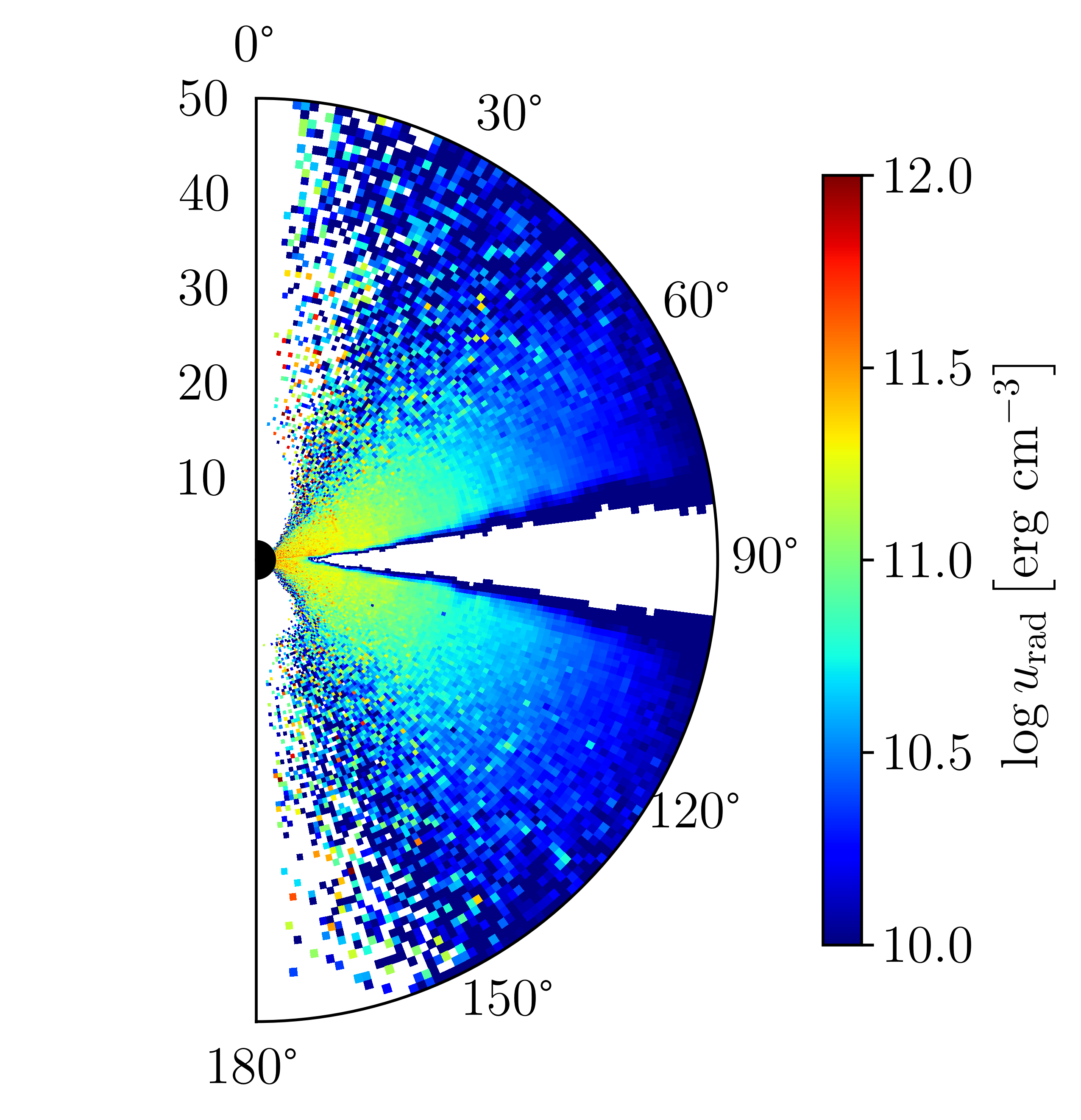}
\caption{Azimuthally-averaged radiation energy density: left, from \textsc{harm3d}'s calculation according to equation \ref{eq:u_rad}; right, the value ``backed out'' from \textsc{pandurata} output via equation \ref{eq:L_IC}. The white polar regions indicate $(r, \theta)$ for which there were no scattering events during \textsc{pandurata}'s Monte Carlo post-processing. The white region near the midplane is the disk body. \label{fig:urads}}
\end{figure}

First, we examine \textsc{harm3d}'s value for the radiation energy density in the corona, equation \ref{eq:u_rad}, compared to \textsc{pandurata}'s. These are shown in Figure \ref{fig:urads}. \textsc{pandurata}'s values are in fact calculated by solving the equation for $L_\mathrm{IC}$ we derived above for use in \textsc{harm3d} (equation \ref{eq:L_IC}) for $u_\mathrm{rad}$, using the supplied $\rho$ values and \textsc{pandurata}'s values for $L_\mathrm{IC}$ consistent with its own solution for $T_e$; \textsc{pandurata} does not explicitly calculate $u_\mathrm{rad}$. Immediately apparent in Figure \ref{fig:urads} (right) is the poor Monte Carlo sampling in the low density region near the poles. Even though a sufficient number of photon packets are launched (1620 from each photosphere surface element) to ensure that the distant observer spectrum is well resolved, scattering events are simply so unlikely in the jet region that it is difficult to evaluate $L_\mathrm{IC}$ there with \textsc{pandurata}; therefore, \textsc{pandurata} values for $u_\mathrm{rad}$ and $T_e$ are unreliable in this region as well. In the much better sampled regions of the corona, it is apparent that \textsc{pandurata}'s $u_\mathrm{rad}$ values agree fairly well with \textsc{harm3d}'s.

\begin{figure}
\plottwo{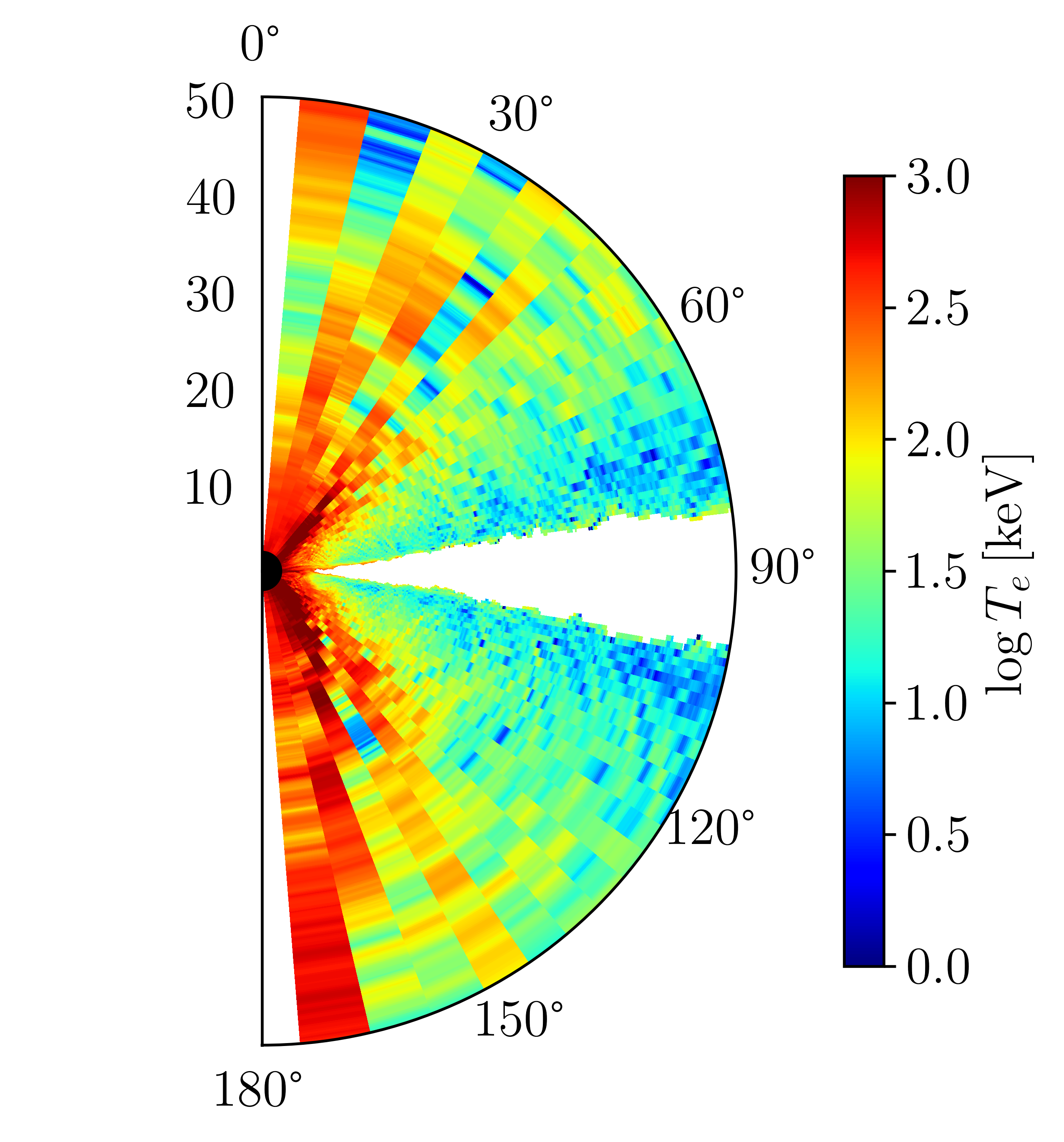}{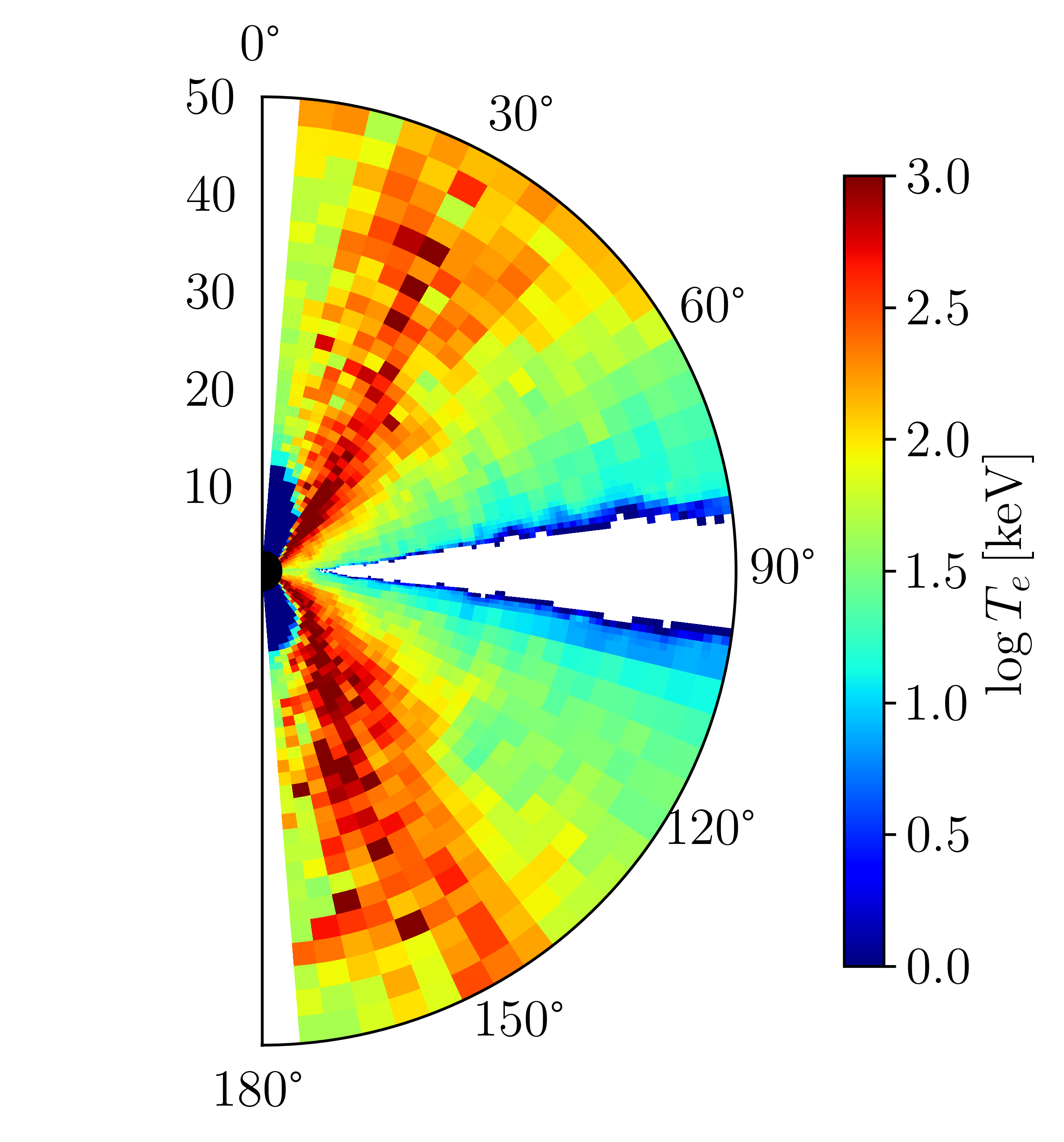}
\caption{Azimuthally-averaged electron temperature: left, from \textsc{harm3d}; right, \textsc{pandurata}'s equilibrium value. The white region near the midplane is the disk body. \label{fig:Tes}}
\end{figure}

\begin{figure}
\plottwo{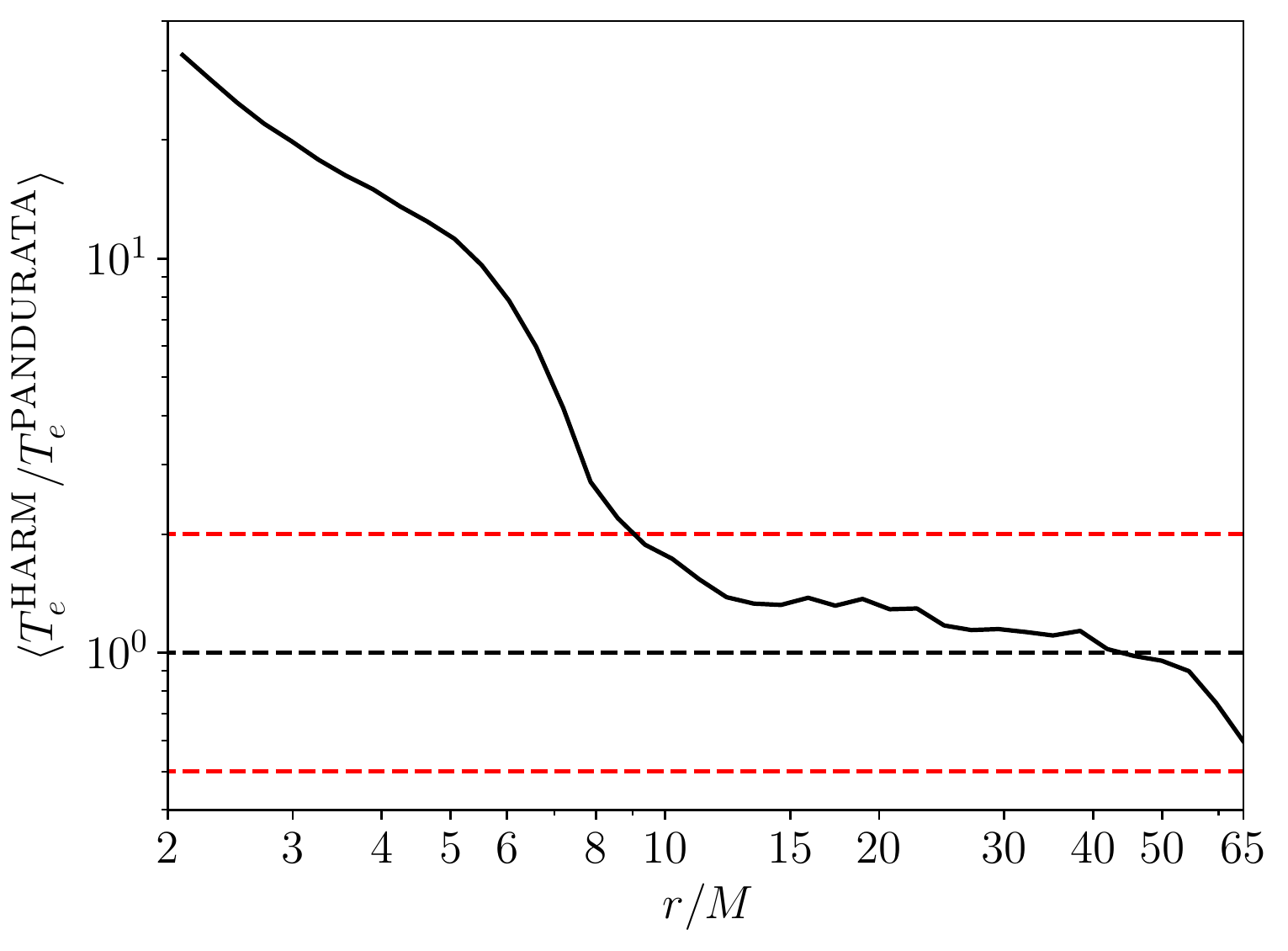}{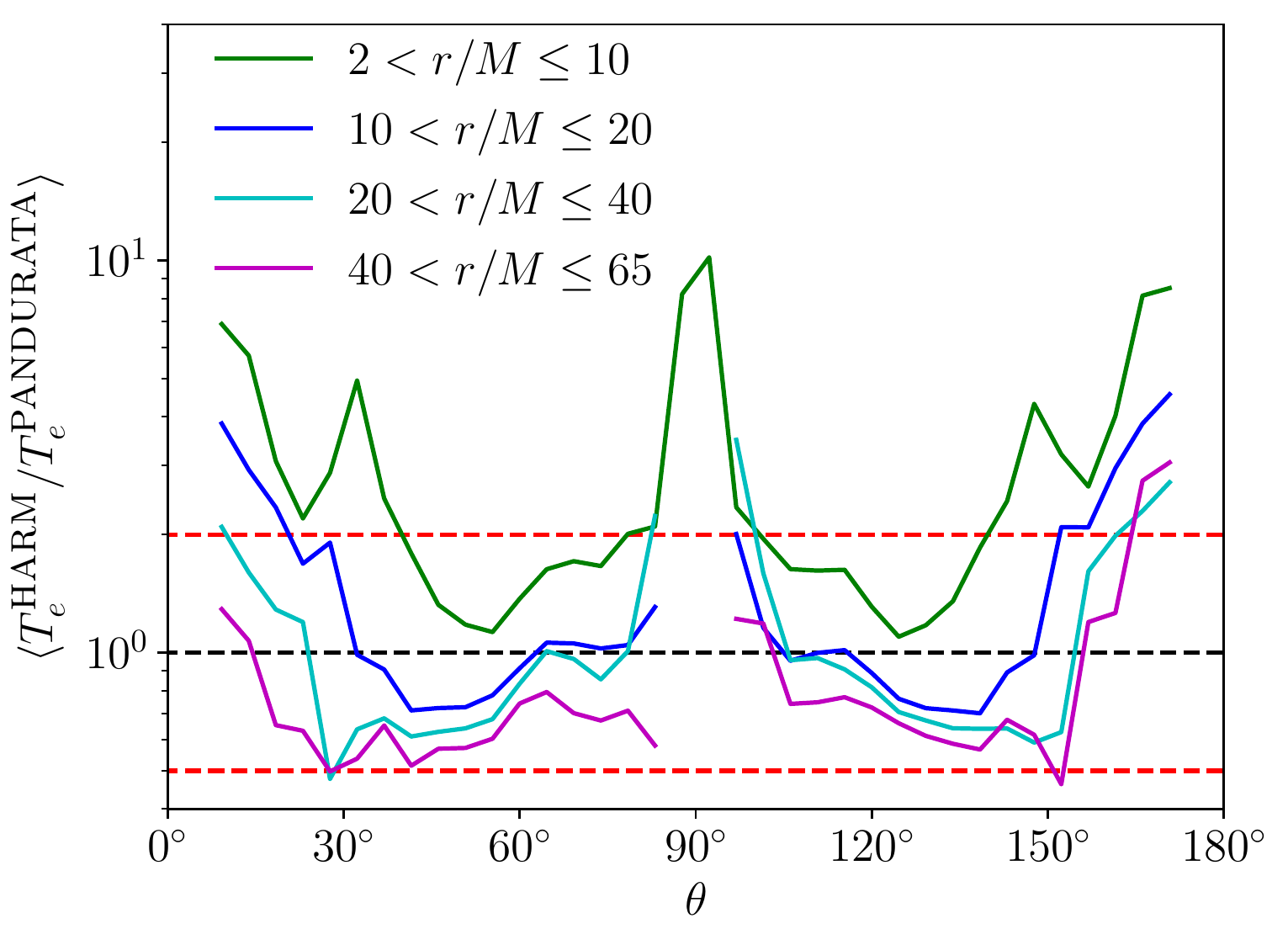}
\caption{The ratio of \textsc{harm3d}'s $T_e$ value to \textsc{pandurata}'s equilibrium $T_e$ value, averaged with density weighting over the full $2000 M$ run. Left: also averaged over polar angle and azimuth, showing dependence of ratio on radius. Right: averaged over radius and azimuth for four contiguous annuli, showing dependence on polar angle. The dashed red lines indicate the boundaries between which $1/2 < \langle T_e^\textsc{harm} / T_e^\textsc{pandurata} \rangle < 2$. \label{fig:Te_errors}}
\end{figure}

\begin{figure}
\epsscale{0.6}
\plotone{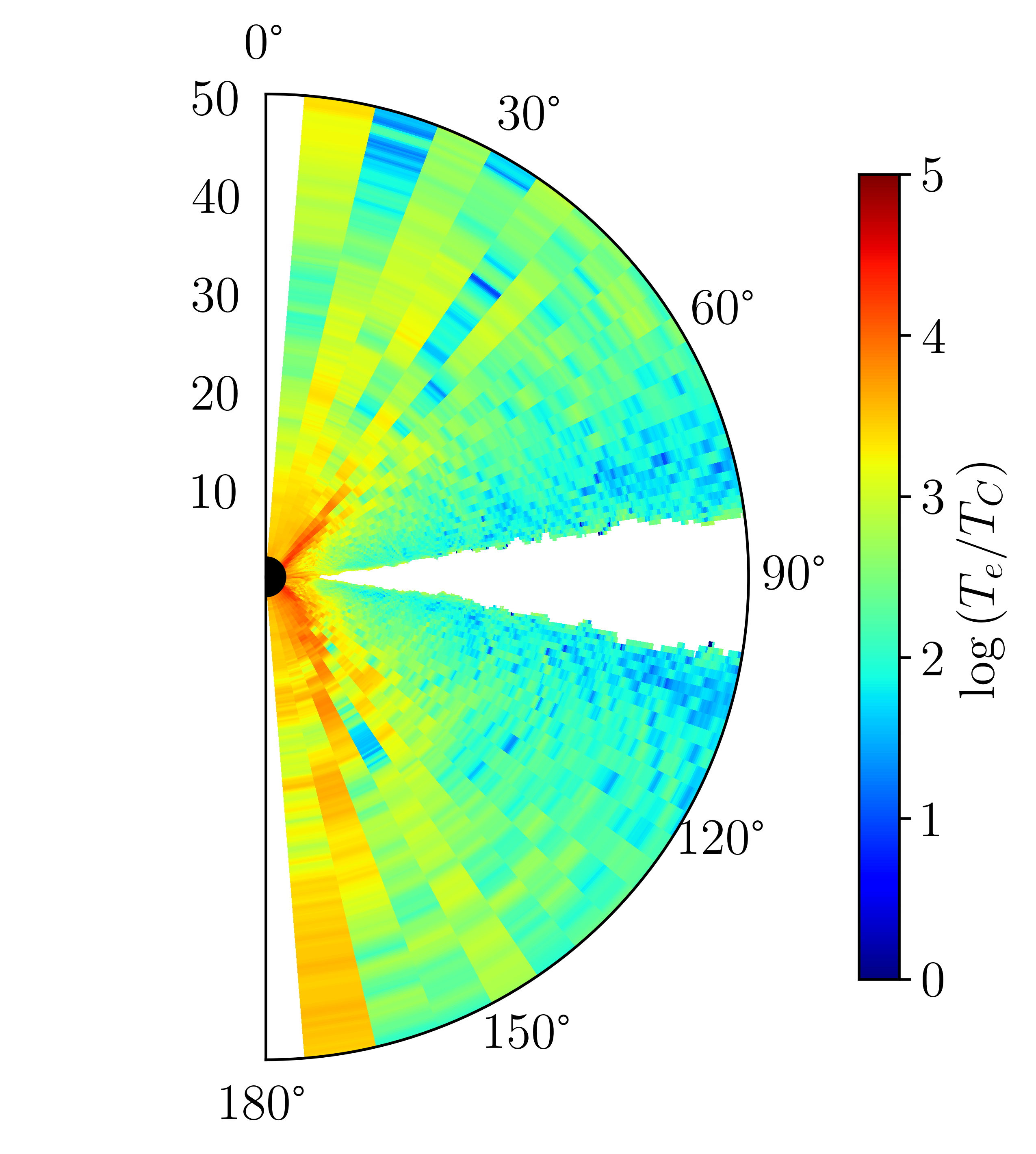}
\caption{Azimuthally-averaged ratio of the \textsc{harm3d}-calculated values for the electron temperature to the Compton temperature. Where the bulk of the luminosity occurs, $T_e/T_C \sim 100$. \label{fig:Te_o_Tc}}
\end{figure}

In Figure \ref{fig:Tes}, we show the azimuthally-averaged \textsc{harm3d} values for $T_e$ compared to \textsc{pandurata}'s equilibrium $T_e$ values for the snapshot at $t = 1000M$. In Figure \ref{fig:Te_errors}, the comparison is presented as the ratio of \textsc{harm3d}'s $T_e$ values to \textsc{pandurata}'s, as a function of radius (left) and polar angle (right), averaged, weighted by density, over the full $2000 M$ run of the simulation. In regions with poor statistical sampling, \textsc{pandurata} will not adjust the temperature from its initial guess---$T_C$ as found by \textsc{harm3d}---which is in effect a lower bound. Figure \ref{fig:Te_o_Tc} shows the azimuthally-averaged ratio of the electron temperature to the Compton temperature. As expected (and required), $T_e \gg T_C$. Thus in Figure \ref{fig:Te_errors} (right), the $T_e$ ratios tend to be very high within $30^\circ$ of the $z$-axis. For much of the (well sampled) coronal volume, however, the two values for $T_e$ are within a factor of 2 of each other. \textsc{harm3d}'s lower values for $u_\mathrm{rad}$ (and correspondingly higher $T_e$) in the corona immediately above and below the inner disk are consistent with ignoring relativity: the relativistic inner disk orbital speeds will preferentially beam seed photons at angles more nearly parallel to the disk surface; in addition, the curved photon trajectories amplify $u_\mathrm{rad}$ near the disk surface in the close vicinity of the black hole. At large radii where relativistic effects are less important, ignoring the occlusion of intervening corona material enhances $u_\mathrm{rad}$ (lowering $T_e$) as compared to \textsc{pandurata}'s value.

\begin{figure}
\epsscale{0.85}
\plotone{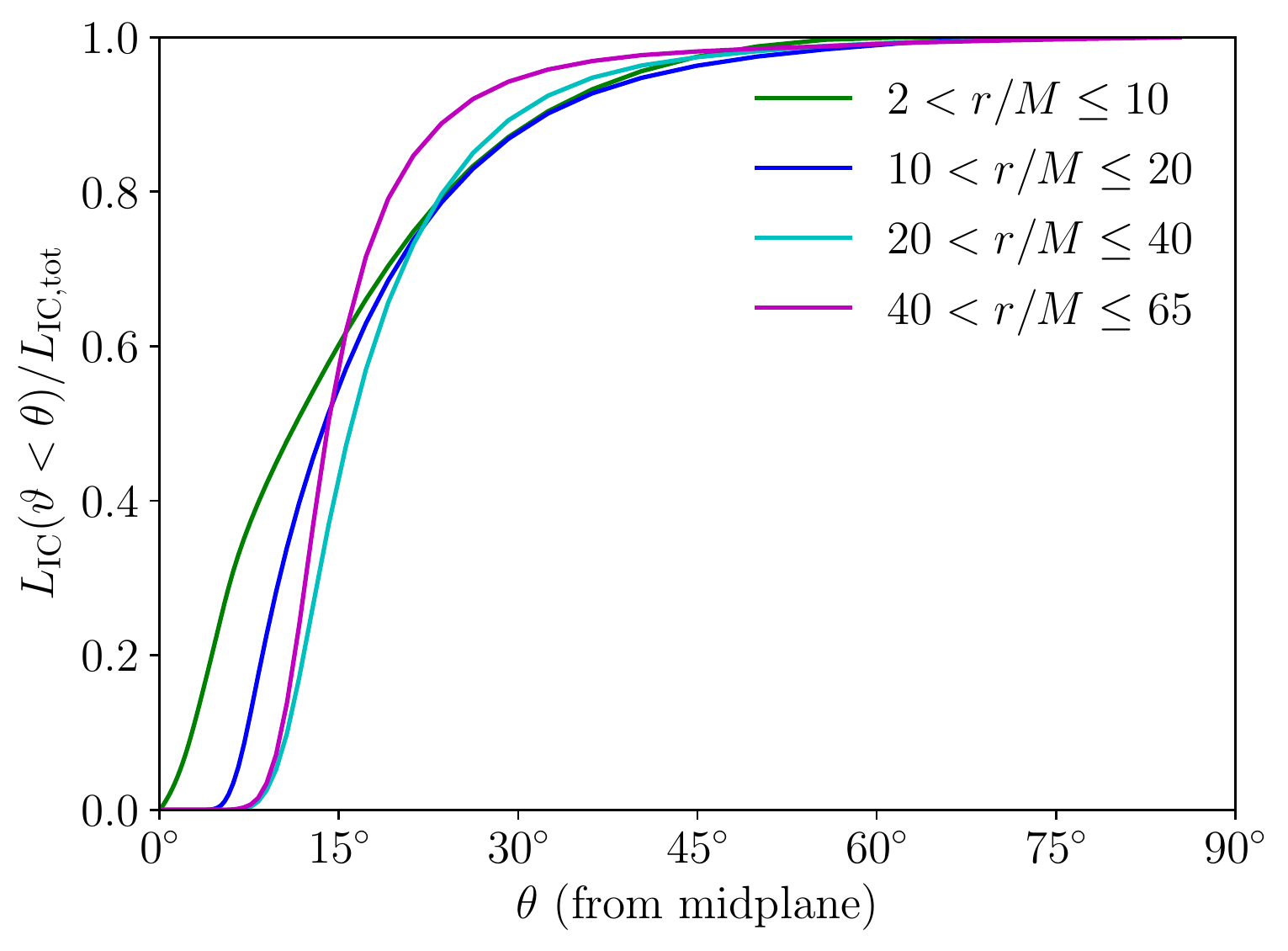}
\caption{The fraction of total corona cooling, for four contiguous annuli, as a function of polar angle as measured from the midplane; time-averaged. \label{fig:L_within_theta}}
\end{figure}

Figure \ref{fig:L_within_theta} shows the fraction of total coronal cooling which occurs within a certain polar angle from the midplane for four contiguous annuli of the simulation volume, averaged over time. We see that the majority of the cooling occurs within $45^\circ$ of the midplane. And as shown in Figure \ref{fig:dLdr_comp}, there is little cooling within $r \simeq 6 M$. Comparing the spatial distribution of corona cooling to the comparison of $T_e$ values in Figure \ref{fig:Te_errors}, we find that in the region which accounts for the majority of the total cooling, \textsc{harm3d}'s estimation of $T_e$ agrees fairly well---at the 10--20\% level---with \textsc{pandurata}'s. From this we conclude that, on net, the procedure detailed in the previous section for calculating the IC cooling rate is generally consistent with a more detailed ray-tracing radiation transport (though still fast light) approach, at a fraction of the cost.

\pagebreak

\section{The Effect of the Inverse Compton Cooling Function on the Observed Spectrum}

\begin{figure}
\epsscale{0.85}
\plotone{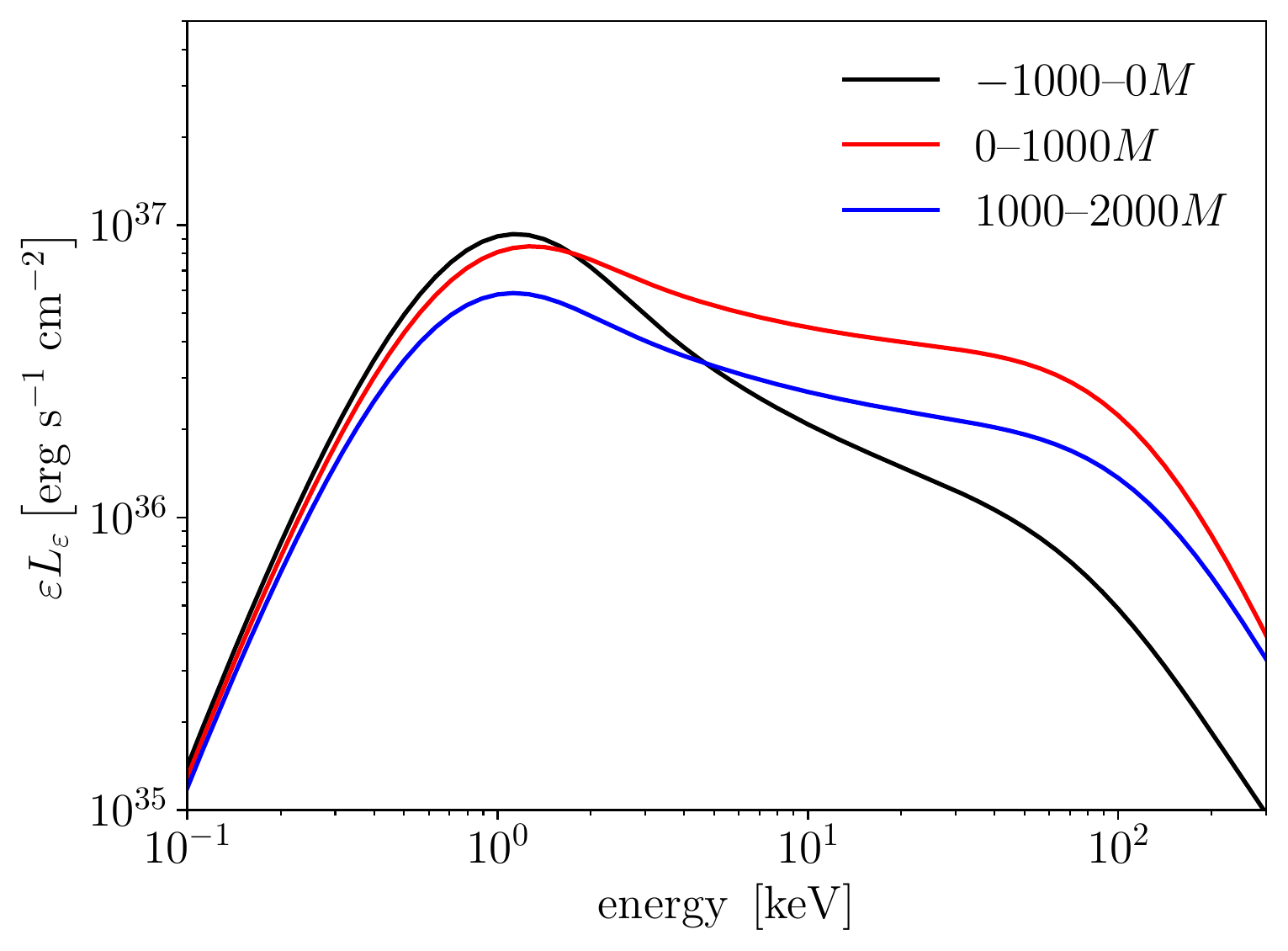}
\caption{The spectral luminosity as seen by a distant observer, averaged over three time intervals. \label{fig:dist_obs_spec}}
\end{figure}

\begin{figure}
\epsscale{0.85}
\plotone{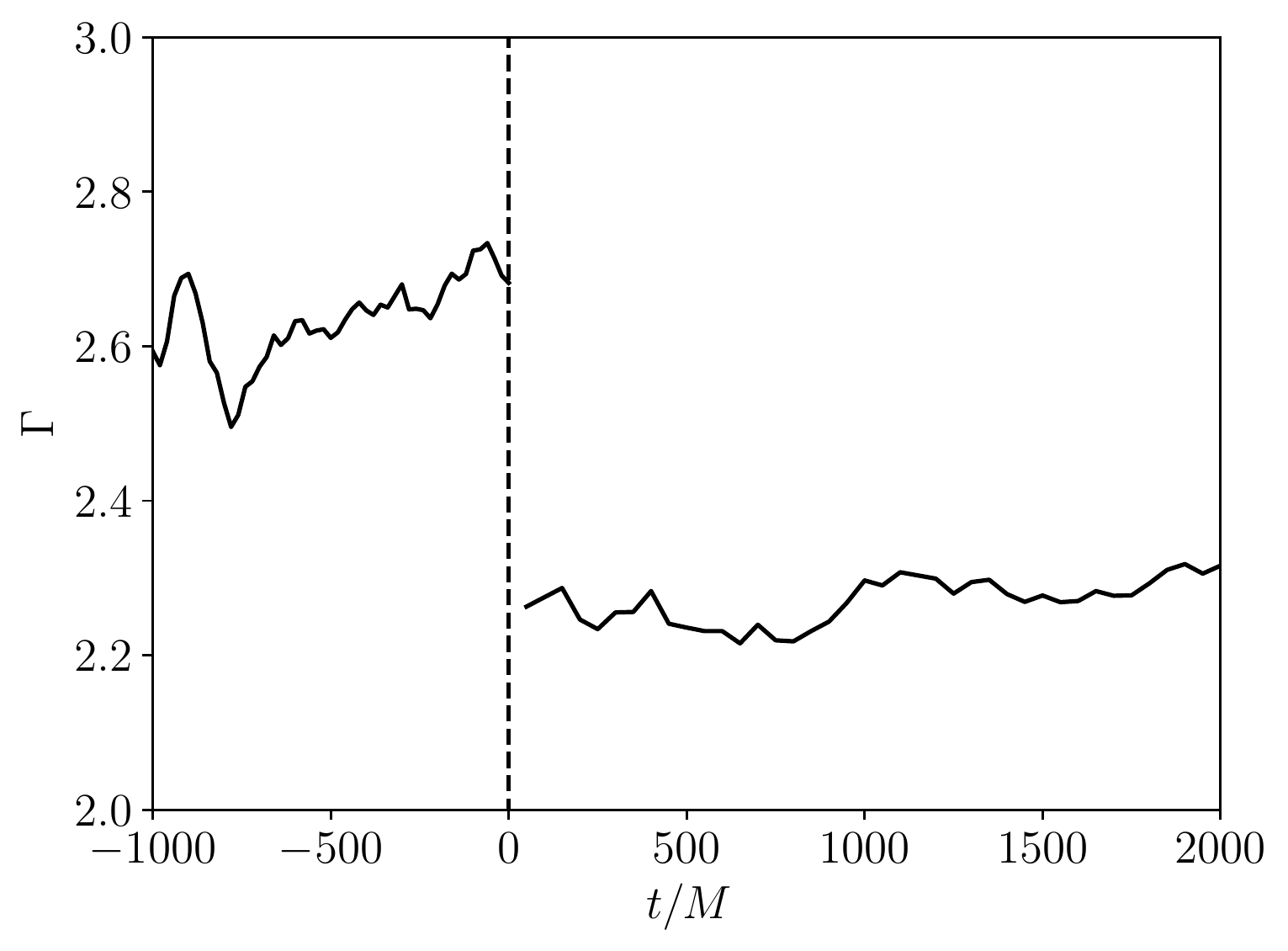}
\caption{The photon index $\Gamma$, fit to the range 2--30 keV of the spectral luminosity, as a function of time. \label{fig:gammas}}
\end{figure}

Following the procedure laid out in \citet{sch13a} and \citet{kin16a, kin19a}, we generate energy- and inclination-dependent simulated spectra by counting those photon packets in \textsc{pandurata} that escape to infinity. The spectral luminosity (integrated over inclination angle) as would be seen by distant observers, averaged over time, is shown in Figure \ref{fig:dist_obs_spec} for the last $1000 M$ of the starter target-temperature simulation and for the first and second $1000 M$ intervals of the simulation with IC cooling function in place. Note that \textsc{pandurata} is applied to multiple snapshots (20 evenly spaced in time) in each interval separately---averaging is performed on the output spectrum, \emph{not} the underlying simulation data. The features of the predicted spectrum are qualitatively similar between the three intervals: a broad thermal peak centered at $\simeq 1$ keV connecting to a power-law tail which rolls over $\gtrsim 60$ keV. However, with the IC cooling function in place, a substantially harder power-law component is achieved. Figure \ref{fig:gammas} shows the variation with time of the photon index $\Gamma$ ($L_\varepsilon/\varepsilon \propto \varepsilon^{-\Gamma}$), measured in the range 2--30 keV. Note that while the bolometric luminosity declines in the final interval, $\Gamma$ is relatively unchanged.

\begin{figure}
\epsscale{0.85}
\plotone{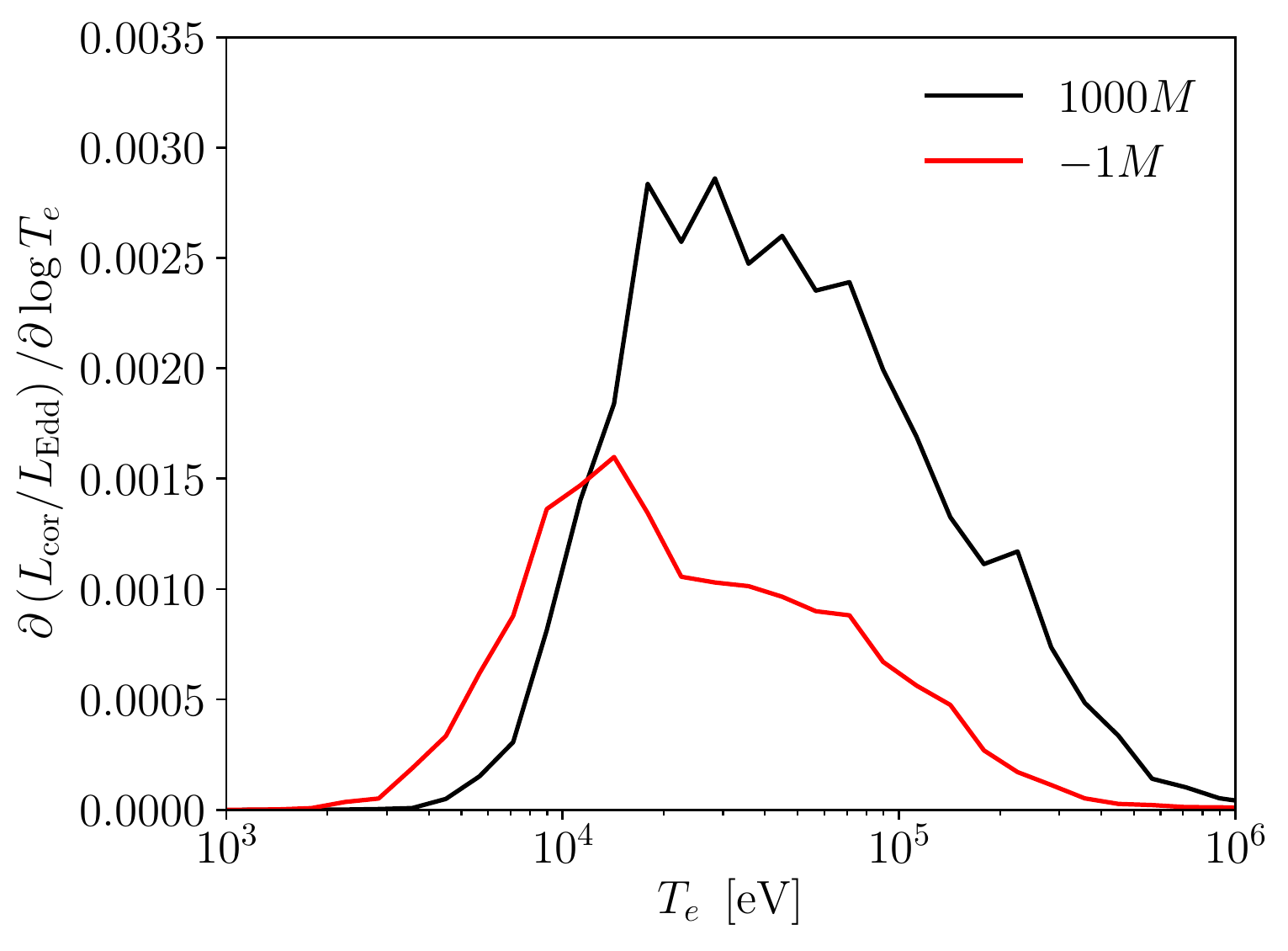}
\caption{The distribution of coronal cooling with respect to its \textsc{pandurata}-determined electron temperature. Two snapshots are shown: just before the IC cooling function is turned on (red) and $1000 M$ later (black). \label{fig:dLdT}}
\end{figure}

The spectral differences between the target-temperature and IC cooling function simulations are explained by the distribution in temperature---as calculated by \textsc{pandurata}---of the cooling corona gas. Figure \ref{fig:dLdT} shows, for the last snapshot of the target-temperature simulation and a representative snapshot of the IC cooling function data, the spread in the \textsc{pandurata}-determined temperature of the cooling gas. In both cases, the coronal gas radiates over a broad range in temperature---however, the relative distribution with temperature is very different. At least for this choice of parameters, the ad hoc target-temperature cooling function posited a cooling rate that was unphysically low for the hot gas, and unphysically high for the cool gas. The result is a softer power-law component in the X-ray spectrum.

\begin{figure}
\epsscale{0.85}
\plotone{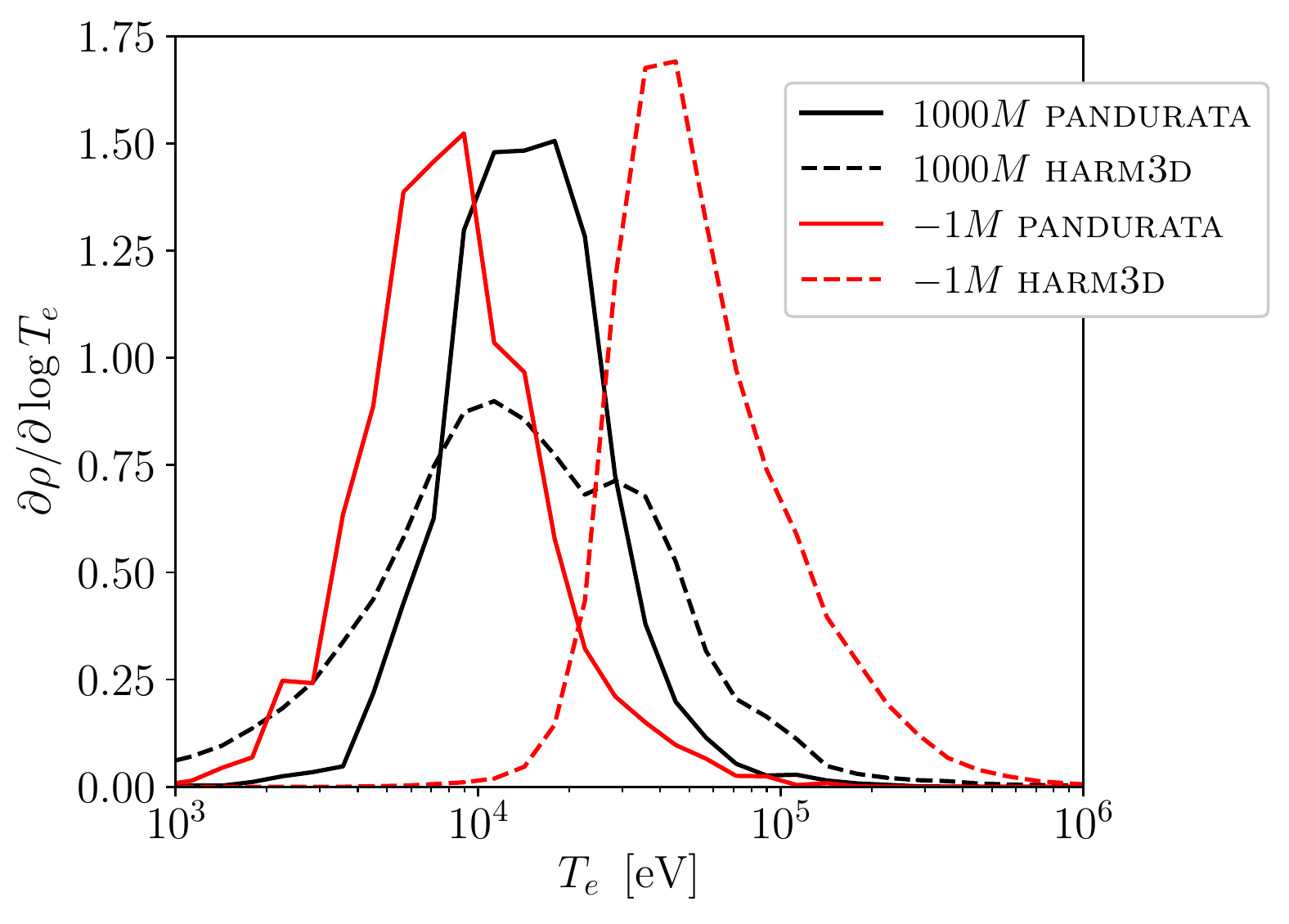}
\caption{The distribution of coronal mass with respect to electron temperature; $T_e$ is determined by \textsc{pandurata} for the dashed lines and by \textsc{harm3d} for solid lines. Two snapshots are shown: just before the IC cooling function is turned on (red) and $1000 M$ later (black). \label{fig:dMdT_comp}}
\end{figure}

Figure \ref{fig:dMdT_comp} shows the coronal mass distribution by temperature for the $-1M$ and $1000M$ snapshots, with $T_e$ as calculated by \textsc{harm3d} (dashed lines) and $T_e$ as calculated by \textsc{pandurata} (solid lines). Note that the agreement between the \textsc{harm3d} and \textsc{pandurata} distributions is substantially improved for the $1000M$ snapshot, for which \textsc{harm3d} uses the IC cooling function. While the \textsc{harm3d}-determined mass-by-$T_e$ distribution has a broader spread in temperature than its \textsc{pandurata}-determined counterpart for the $t = 1000M$ snapshot, their peaks---i.e., where most of the material lies---agree. By its nature as a Monte Carlo code, \textsc{pandurata} will best sample the most dense region just outside the disk (recall Figure \ref{fig:urads}, right); the high altitude regions where \textsc{harm3d}- and \textsc{pandurata}-determined temperatures most differ (Figure \ref{fig:Te_errors}, right), and where only a small fraction of the total cooling takes place (Figure \ref{fig:L_within_theta}), account for the wider breadth the \textsc{harm3d} distribution.

\pagebreak

\section{Two Temperature Inverse Compton Cooling Function}
\label{2T_section}

Below we describe our procedure for relaxing the assumption that $T_e = T_i$ in a particular scenario:

\begin{enumerate}

\item The ion and electron populations are individually in thermal equilibrium locally, described by $T_i$ and $T_e$ at each point in the corona.

\item Turbulent energy is dissipated into the ions \emph{only}.

\item Energy is exchanged between the ion and electron populations through Coulomb collisions.

\item $T_e$ adjusts instantaneously so that the rate at which energy is \emph{added} to the electron population---through either Coulomb collisions or Compton heating---is equal to the rate at which energy is \emph{lost} due to inverse Compton cooling.

\end{enumerate}

The problem is to find $T_e$ and $T_i$ given the above assumptions.

We begin with the relativistically correct ion-electron energy exchange rate derived in \citet{ste83a, ste83b}:

\begin{eqnarray}
\frac{d}{dt} u_e\ \mathrm{(Coulomb\ heating)} = \frac{3}{2} \frac{m_e}{m_i} \sigma_T c \ln \Lambda n_e n_i \left( k_B T_i  - k_B T_e \right) \label{eq:coulomb_heating} \\
\times \left\{ \frac{1}{K_2(1/\Theta_e) K_2(1/\Theta_i)} \left[ \frac{2 (\Theta_e + \Theta_i)^2 + 1}{\Theta_e + \Theta_i} K_1 \left( \frac{\Theta_e + \Theta_i}{\Theta_e \Theta_i} \right) + 2 K_0 \left( \frac{\Theta_e + \Theta_i}{\Theta_e \Theta_i} \right) \right] \right\} \nonumber \\
= \frac{3}{2} \frac{m_e}{m_i} \sigma_T c \ln \Lambda n_e n_i \left( k T_i  - k T_e \right) f(\Theta_e, \Theta_i), \nonumber
\end{eqnarray}

\noindent where $u_e$ is the internal energy per unit volume of the electron population, $\ln \Lambda$ is the Coulomb logarithm (the logarithm of the ratio of the maximum to minimum impact parameters; $\ln \Lambda \sim 20$), $\Theta_i$ is the dimensionless ion temperature, equal to $k_B T_i/m_i c^2$, and $K_n$ is the $n^\mathrm{th}$ order modified Bessel function of the second kind. The term in braces constitutes $f(\Theta_e, \Theta_i)$.

Electrons gain or lose energy to ions through Coulomb collisions (equation \ref{eq:coulomb_heating}), gain energy from photons through Compton heating (equation \ref{eq:comp_heat}) and lose energy to photons through Compton cooling (equation \ref{eq:L_IC}). The equilibrium $T_e$ is that for which these processes balance, $du_e/dt = 0$. Setting the sum of these three processes to zero, we have:

\begin{equation}
\frac{3}{2} \frac{m_e}{m_i} \sigma_T c \ln \Lambda n_e n_i \left( k_B T_i  - k_B T_e \right) f(\Theta_e, \Theta_i) + \frac{\sigma_T}{m_e c} n_e u_\mathrm{rad} \langle \varepsilon \rangle - 4 \sigma_T c n_e u_\mathrm{rad} \Theta_e (1 + 4 \Theta_e) = 0.
\end{equation}

\noindent With the identification that $n_e = \chi n_i$ and $n_i = \rho/m_i$, and some algebraic manipulation, we rewrite the above as:

\begin{equation}
\frac{3}{2} \frac{m_e}{m_i} \ln \Lambda \left( \Theta_i - \frac{m_e}{m_i} \Theta_e \right) f( \Theta_e, \Theta_i ) + \frac{u_\mathrm{rad}}{\rho c^2} \frac{ \langle \varepsilon \rangle }{m_e c^2} - 4 \frac{u_\mathrm{rad}}{\rho c^2} \Theta_e (1 + 4 \Theta_e) = 0.
\label{eq:T_e_balance}
\end{equation}

\noindent We rearrange the ideal gas law, equation \ref{eq:ideal_gas_law}, to solve for $\Theta_i$,

\begin{equation}
\Theta_i = (c_P/c_V - 1) \frac{u}{\rho c^2} - \chi \frac{m_e}{m_i} \Theta_e.
\label{eq:Theta_i}
\end{equation}

\noindent Substituting the above into equation \ref{eq:T_e_balance}:

\begin{eqnarray}
\frac{3}{2} \frac{m_e}{m_i} \ln \Lambda \left[ (c_P/c_V - 1) \frac{u}{\rho c^2} - \frac{m_e}{m_i} (1 + \chi) \Theta_e \right] & f \left( \Theta_e, (c_P/c_V - 1) \frac{u}{\rho c^2} - \chi \frac{m_e}{m_i} \Theta_e \right) \label{eq:T_e_balance_v2} \\
+ & \frac{u_\mathrm{rad}}{\rho c^2} \frac{ \langle \varepsilon \rangle }{m_e c^2} - 4 \frac{u_\mathrm{rad}}{\rho c^2} \Theta_e (1 + 4 \Theta_e) = 0. \nonumber
\end{eqnarray}

\noindent We define three dimensionless quantities,

\begin{equation}
A \equiv \frac{u}{\rho c^2},\ B \equiv \frac{u_\mathrm{rad}}{\rho c^2},\ \mathrm{and}\ C \equiv \frac{\langle \varepsilon \rangle}{m_e c^2},
\end{equation}

\noindent which we substitute into equation \ref{eq:T_e_balance_v2}:

\begin{eqnarray}
\frac{3}{2} \frac{m_e}{m_i} \ln \Lambda \left[ (c_P/c_V - 1) A - \frac{m_e}{m_i} (1 + \chi) \Theta_e \right] & f \left( \Theta_e, (c_P/c_V - 1) A - \chi \frac{m_e}{m_i} \Theta_e \right) \label{eq:T_e_balance_v3} \\
+ & B C - 4 B \Theta_e (1 + 4 \Theta_e) = 0. \nonumber
\end{eqnarray}

We arrive at an equation for $\Theta_e$, in terms of only dimensionless quantities, which depends on three parameters which can be read off directly from already-computed values in \textsc{harm3d}. Equation \ref{eq:T_e_balance_v3} is not amenable to real-time solution in each coronal cell each timestep; rather, we tabulate its solution on a grid covering all possible, reasonable values of $A$, $B$, and $C$, and use trilinear interpolation to calculate the appropriate equilibrium value of $\Theta_e$ from said lookup-table in the course of the simulation run. From $\Theta_e$ we calculate $\Theta_i$ from equation \ref{eq:Theta_i}. The cooling function which enters into \textsc{harm3d} is the net Compton cooling translated to code units.

\begin{figure}
\epsscale{0.85}
\plotone{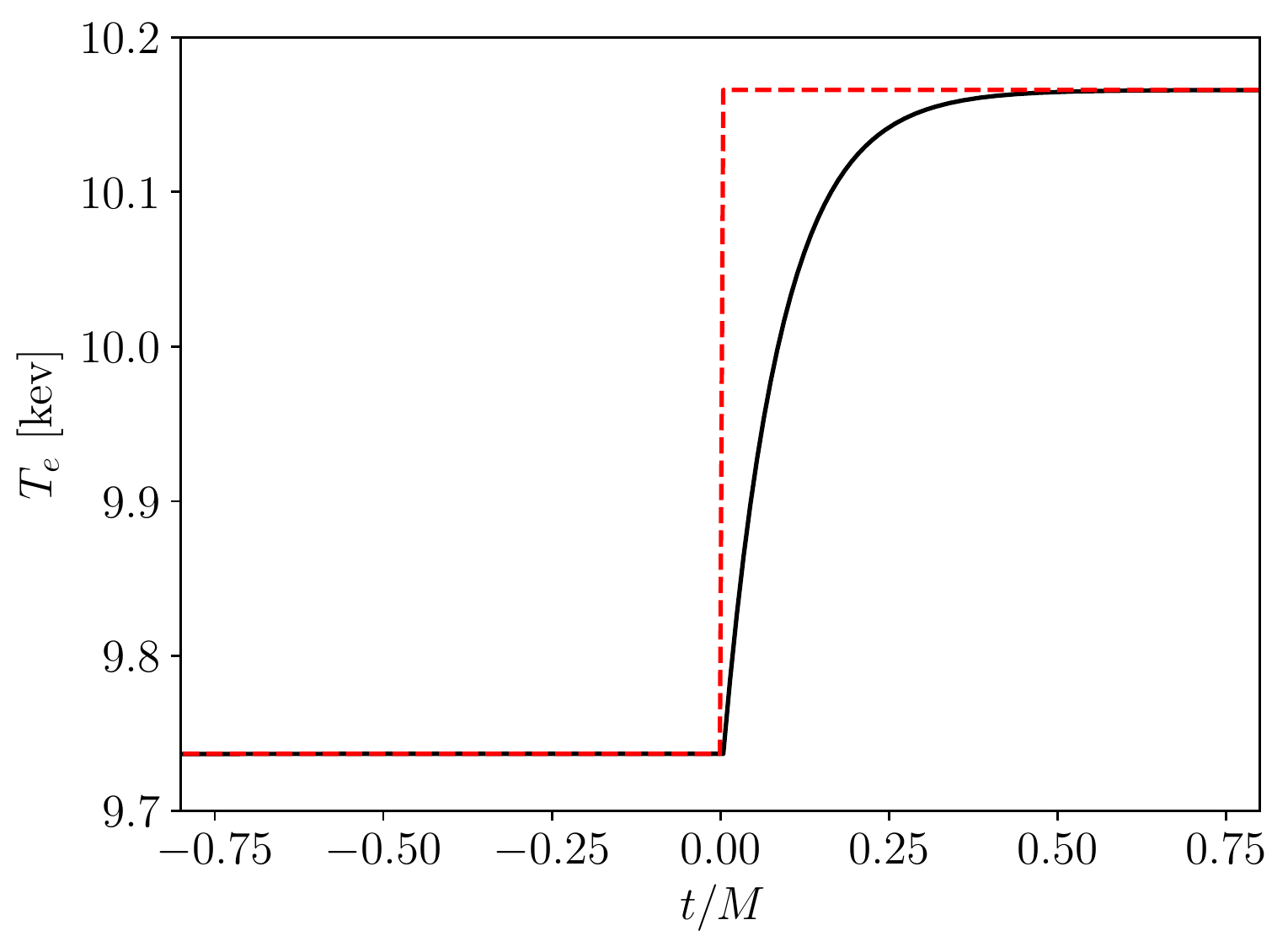}
\caption{An integration for $T_e$ of the Coulomb and Compton rate equations using values typical for the corona, in a scenario where the ion temperature doubles at $t = 0$. The electron temperature adjusts to its new equilibrium value after approximately $0.3 M$. Our model treats this as instantaneous (the red dashed line). \label{fig:Te_int}}
\end{figure}

Figure \ref{fig:Te_int} is a demonstration of our ``snap to equilibrium'' approximation for $T_e$. It shows a simple forward Euler integration of the Compton and Coulomb heating/cooling equations, with $n_e = 3 \times 10^{16}\ \mathrm{cm}^{-3}$ and $u_\mathrm{rad} = 3 \times 10^{12}\ \mathrm{erg\ cm}^{-3}$, values typical for the corona of a $10 M_\odot$ black hole accreting at 1\% Eddington. For $t < 0$, $T_i = 60$ keV. At $t = 0$, the ion temperature jumps to 120 keV. The adjustment of $T_e$ to its new equilibrium value takes $\simeq 0.3 M$. We approximate such an adjustment to be instantaneous. Of course, $n_e$ and $u_\mathrm{rad}$---which dictate this re-equilibration time scale---vary broadly in time and space in the corona. The range of actually encountered values for $n_e$ and $u_\mathrm{rad}$ (in regions where there is any substantial cooling) correspond to adjustment times in the range $0.1$--$1 M$.

\section{Comparison of the 1T and 2T Cooling Functions}

\begin{figure}
\epsscale{0.85}
\plotone{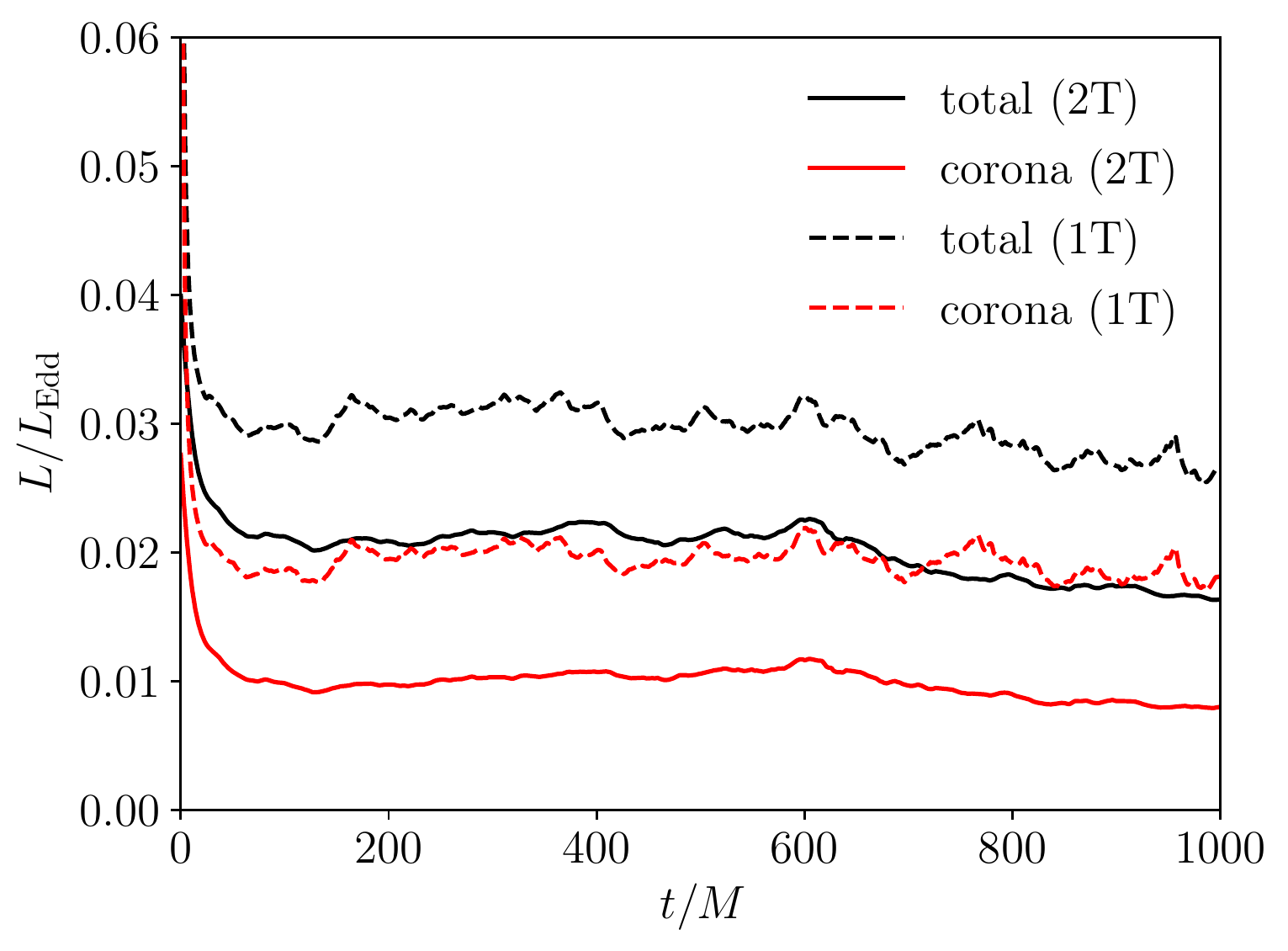}
\caption{The total cooling rate, with the contribution from the corona only, for both 2T and 1T simulations, as fractions of the Eddington luminosity, as functions of time. \label{fig:cooling_2T}}
\end{figure}

\begin{figure}
\epsscale{0.85}
\plotone{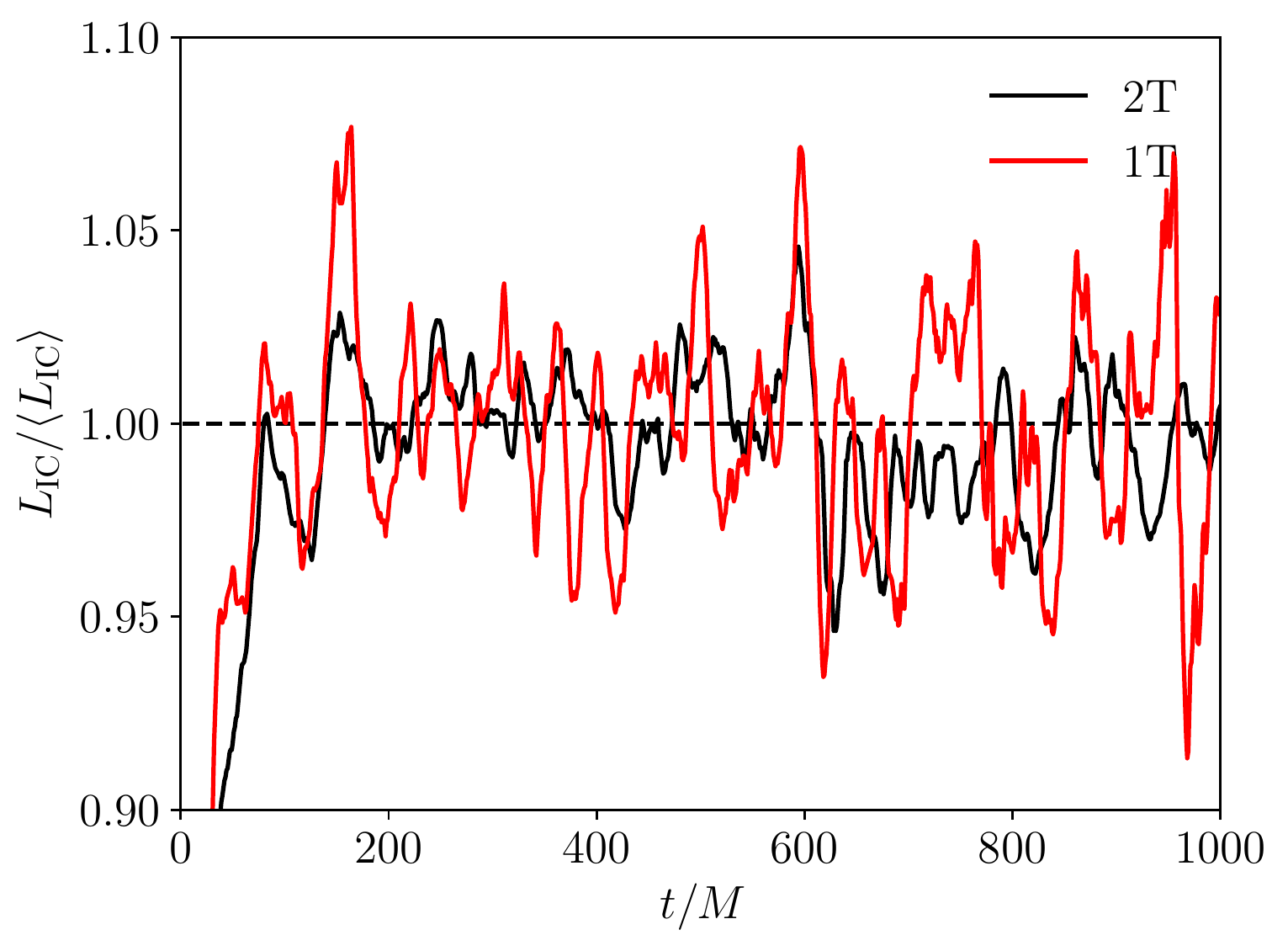}
\caption{The volume-integrated coronal cooling rate, divided by a $25 M$ window moving average of the same data, for both 2T and 1T simulations. \label{fig:cooling_variation}}
\end{figure}

\begin{figure}
\epsscale{0.85}
\plotone{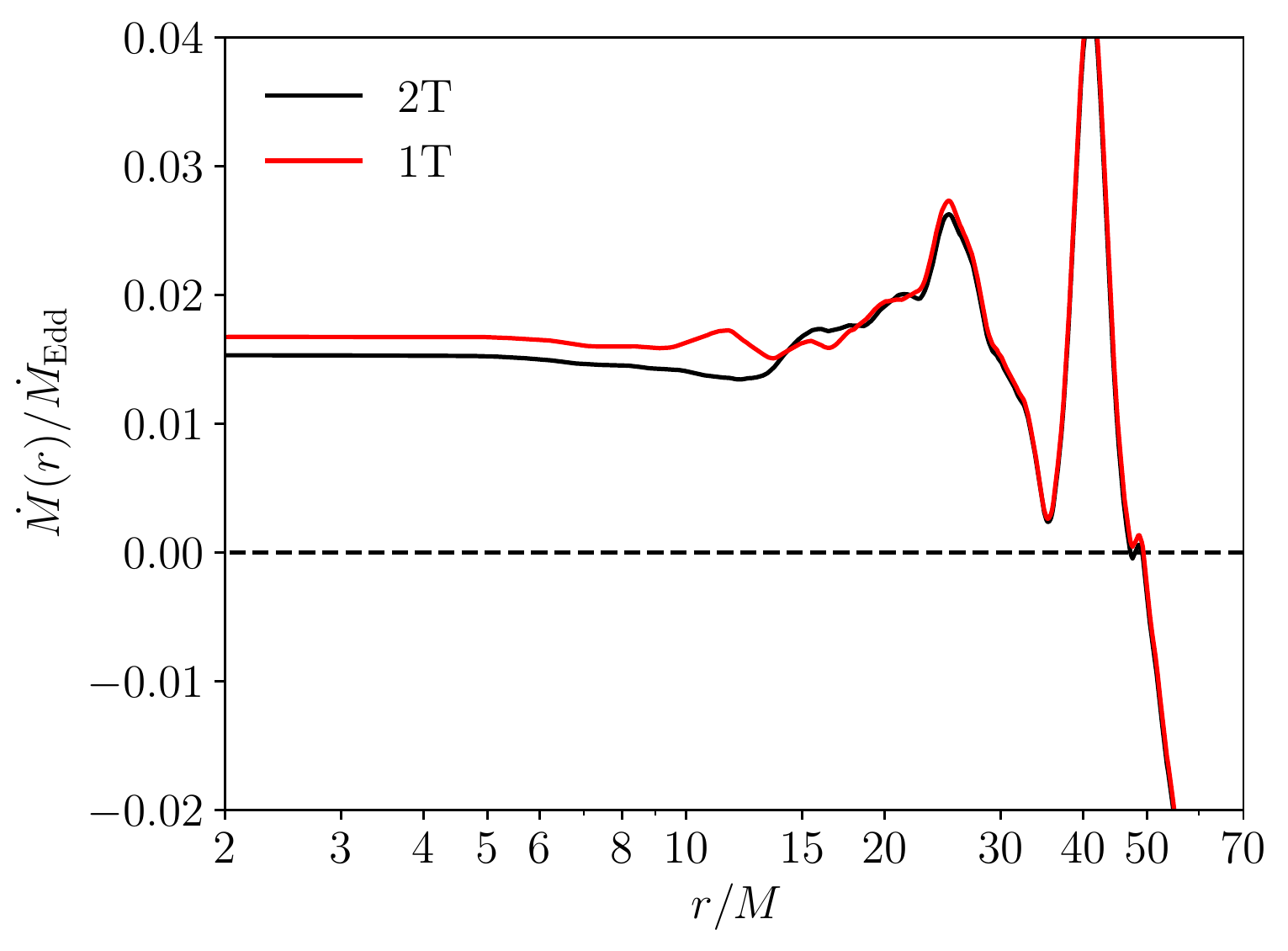}
\caption{For both 2T and 1T simulations: the shell-integrated mass inflow rate, $\dot{M}$, as a function of radial coordinate, averaged in time over the 0--$1000 M$ window, expressed in ratio to the nominal Eddington mass accretion rate. \label{fig:inflow_2T}}
\end{figure}

The two-temperature (2T) IC cooling function is applied to the same starter simulation to which we applied the 1T IC cooling function in the previous section. With the 2T IC cooling function switched on, the system is evolved for $1000 M$. In Figure \ref{fig:cooling_2T} we compare the total and corona-only volume-integrated cooling rates for both approaches. Note that because the IC cooling function is applied only in the corona, the disk component for each approach is not shown as they are nearly identical. Not surprisingly, the 2T method results in a lower corona luminosity; the mean disk fraction is, as with the old target-temperature cooling function, nearly exactly half. The mass accretion rate at the event horizon is also somewhat lower (Figure \ref{fig:inflow_2T}). Comparing the time-averaged luminosity (the ``ray-traced'' power---not shown in Figure \ref{fig:cooling_2T} for clarity, though as in Figure \ref{fig:cooling}, it is nearly the same as the total volume-integrated cooling rate), we find that the 2T approach yields a radiative efficiency of $\eta = 0.0804$, slightly lower than the value for the 1T approach yet still more efficient than either the target-temperature run or NT. Also, notice that the coronal luminosity for the 2T simulation has lesser short time variability than for the 1T simulation. Figure \ref{fig:cooling_variation} compares the volume-integrated corona power divided by a moving average of itself (over a $25 M$ window). The standard deviation of the moving time-averaged coronal luminosity, $L_\mathrm{IC}/\langle L_\mathrm{IC} \rangle$, is twice as large for the 1T simulation than for the 2T simulation. This is consistent with the demonstration in Figure \ref{fig:Te_int}: in that example, $T_i$ jumped from 60 keV to 120 keV, while the equilibrium $T_e$ value increased by less than 1 keV. Because IC power is a function of the \emph{electron} temperature---not the ion temperature---the energy exchange between ions and electrons dampens variability.

\begin{figure}
\epsscale{0.6}
\plotone{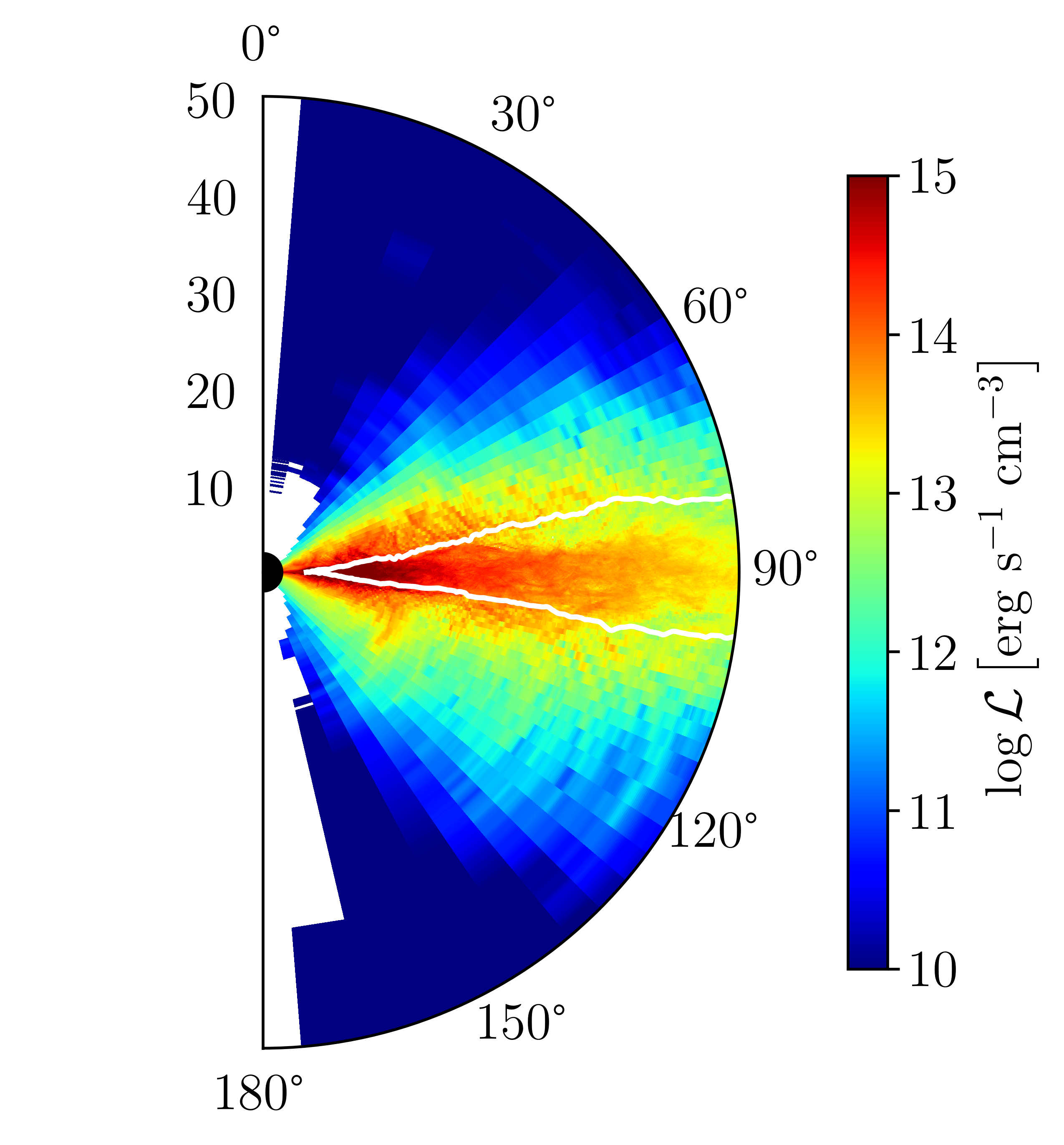}
\caption{Azimuthally-averaged values of the fluid frame cooling rate, for the snapshot at $t = 1000 M$ of the 2T simulation. The white lines indicate the ($\phi$-averaged) photosphere surfaces. Cells with zero cooling are shown in white. \label{fig:cool_2T}}
\end{figure}

\begin{figure}
\plottwo{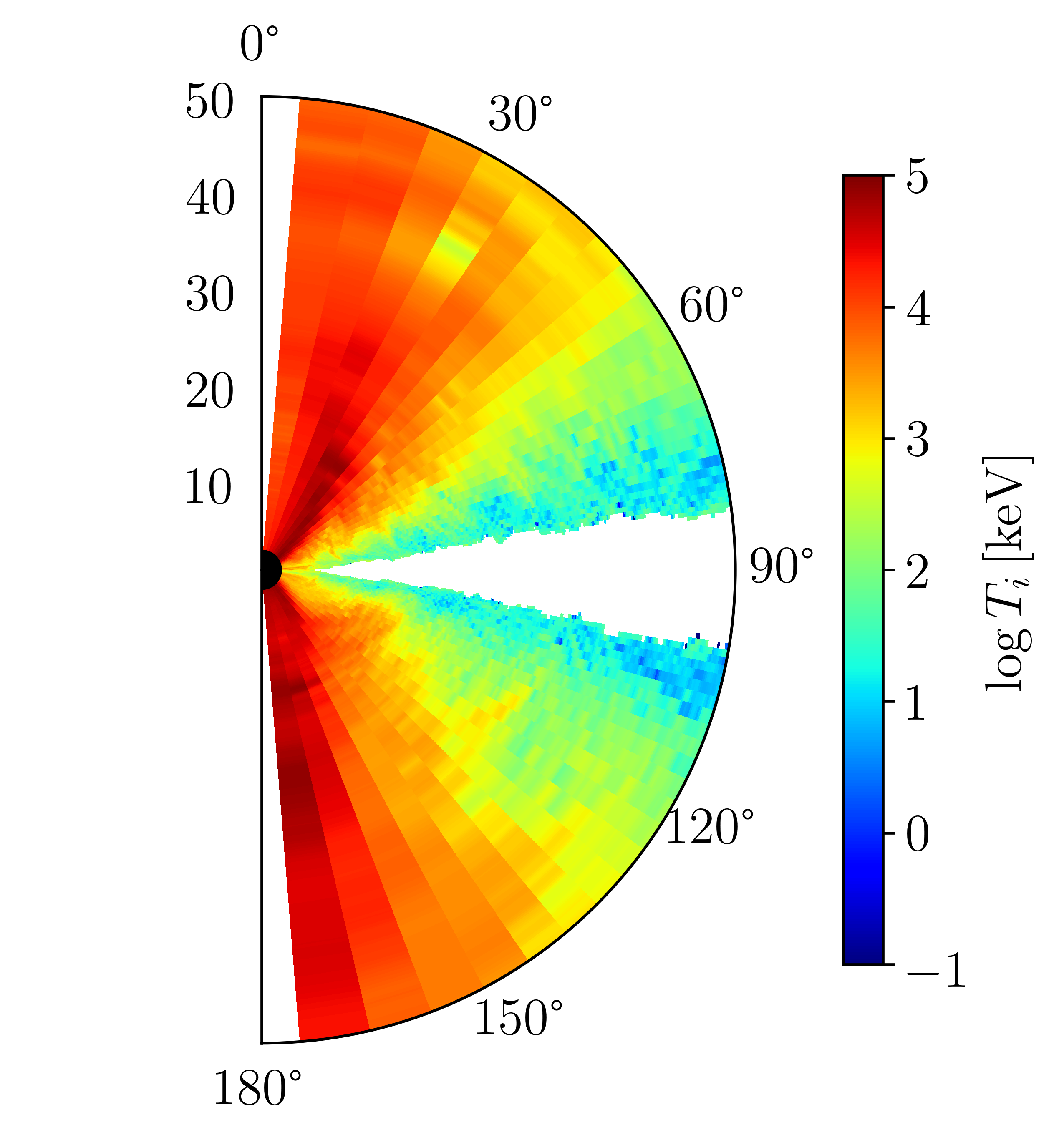}{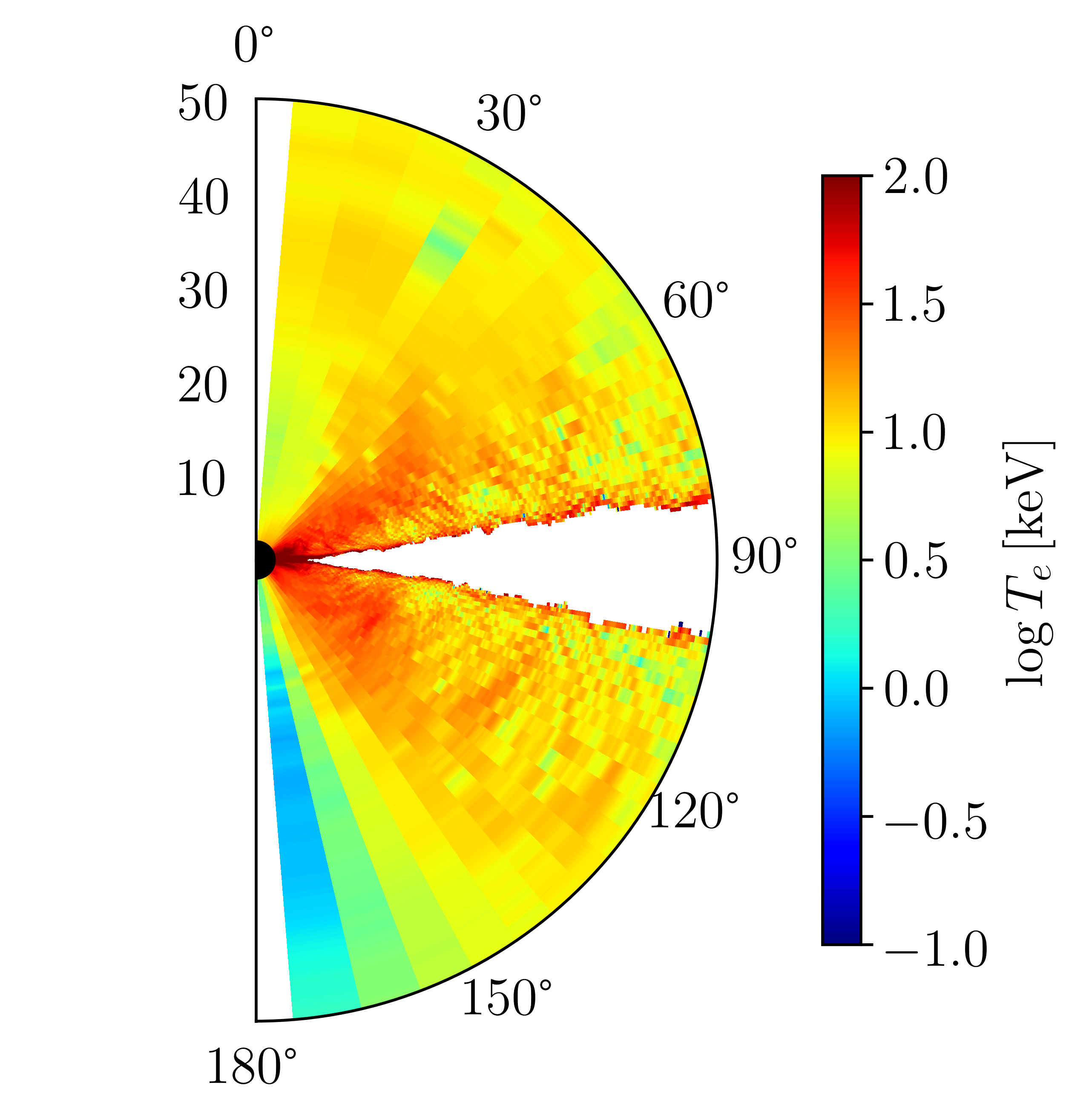}
\caption{Azimuthally-averaged values of the ion temperature (left) and electron temperature (right). Note the dramatic difference in scales between the two plots. The disk body (where no distinction is made) is shown in white. \label{fig:Ti_and_Te}}
\end{figure}

\begin{figure}
\plottwo{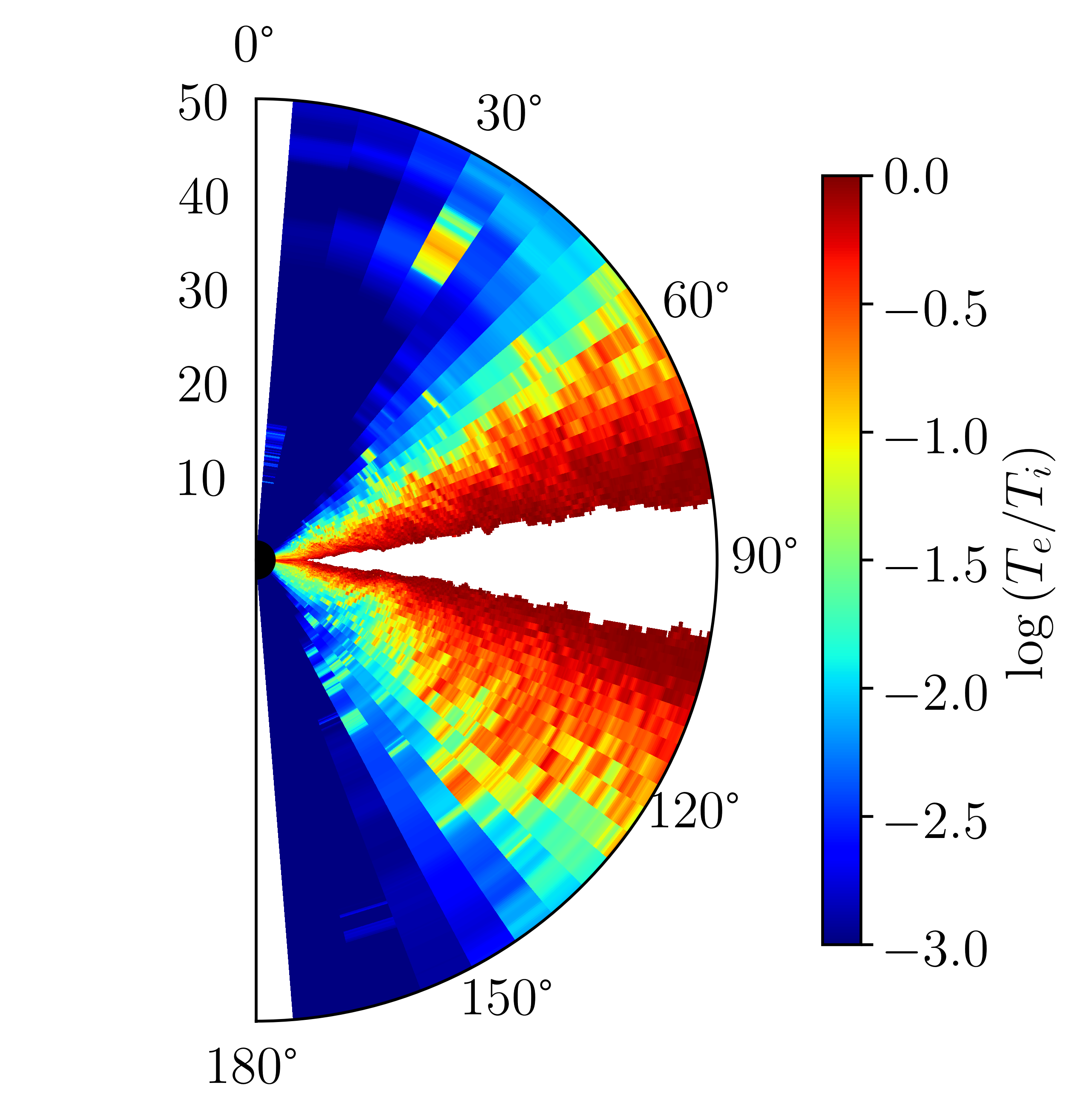}{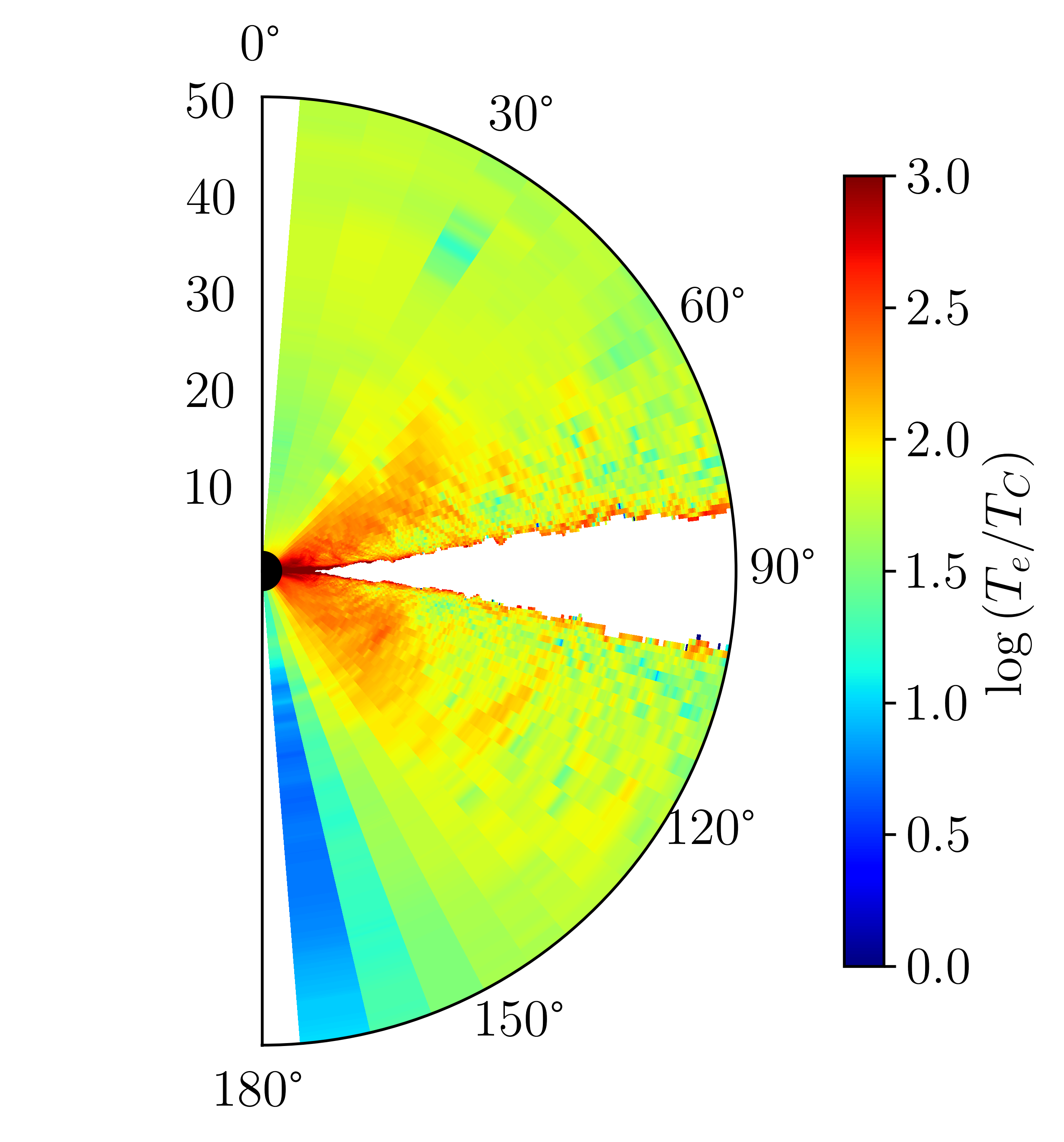}
\caption{Azimuthally-averaged values of the ratio of the electron to ion temperatures (left) and electron to Compton temperatures (right). The disk body is shown in white. \label{fig:Te_o_Tic}}
\end{figure}

Indeed, Coulomb collisions act as a bottleneck between MHD heating and IC cooling. Figure \ref{fig:cool_2T} shows the azimuthally-averaged cooling rate for the 2T simulation at $t = 1000 M$. Compared to the same plot for the 1T simulation (Figure \ref{fig:rho_and_cool}, right), the cooling is confined to a smaller region, closer to the disk. The ion-electron exchange rate is proportional to the square of the density---further from the midplane, the density is simply too low to support Coulomb heating of the electrons. The temperature plots of Figures \ref{fig:Ti_and_Te} and \ref{fig:Te_o_Tic} show this clearly. Compare to the plot of $T_e$ (and, by assumption, $T_i$) for the 1T simulation in Figure \ref{fig:Tes} (left): the ions are hotter and the electrons are cooler. The ratio $T_e/T_i$ declines sharply from near unity just outside the disk photosphere to $< 10^{-3}$ in the jet cone. The decreased ion cooling rate with the 2T method relative to the 1T accounts for the decreased mass accretion rate: greater ion temperature translates to a stronger pressure support against gravity. Note also that while $T_e > T_C$ in the 2T case as well as the 1T, the ratio here is smaller, $T_e/T_C \gtrsim 30$.

\begin{figure}
\epsscale{0.85}
\plotone{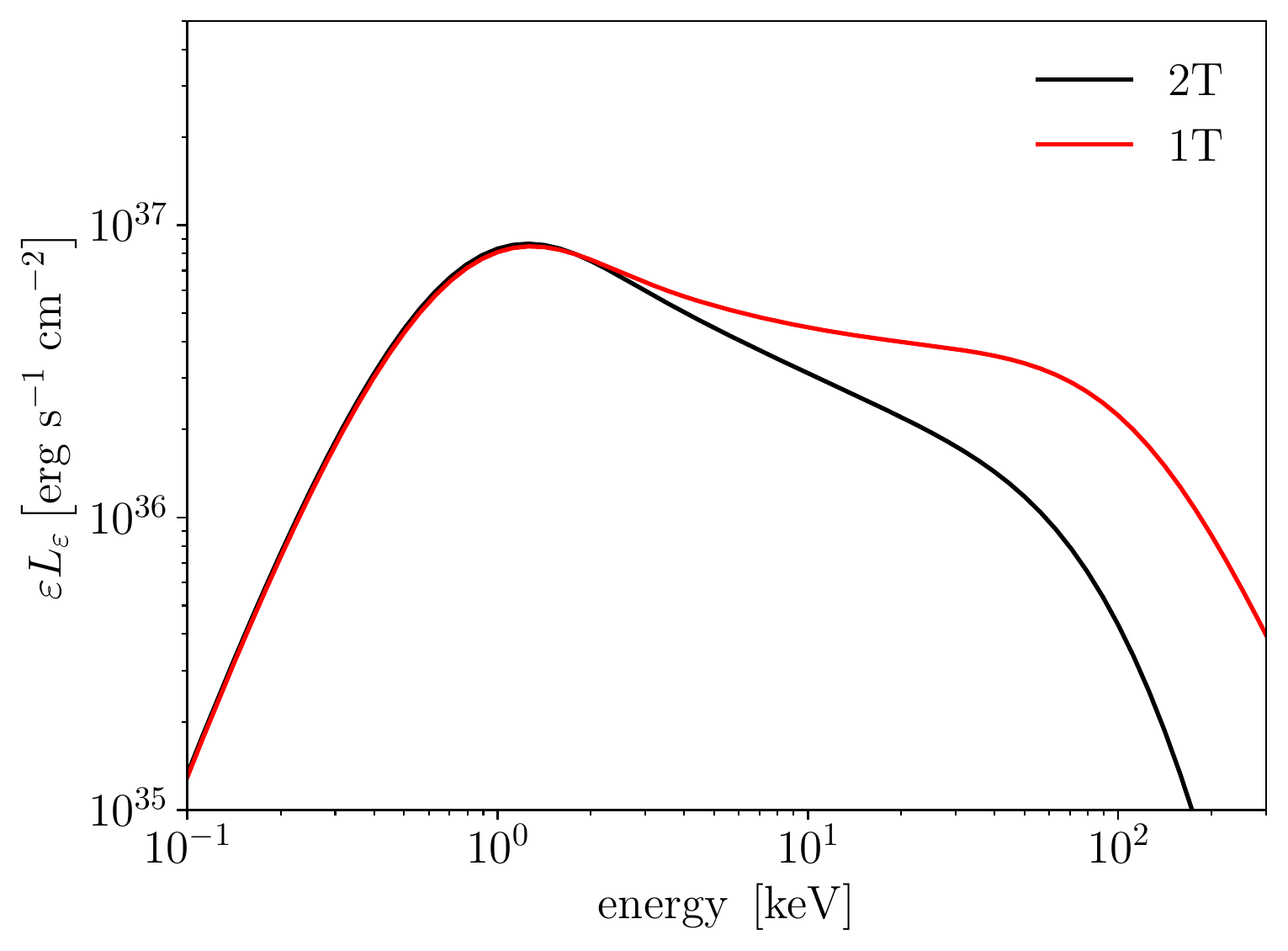}
\caption{The spectral luminosity as seen by a distant observer, averaged over 0--$1000M$ for both 1T and 2T data. \label{fig:dist_obs_spec_2T}}
\end{figure}

\begin{figure}
\epsscale{0.85}
\plotone{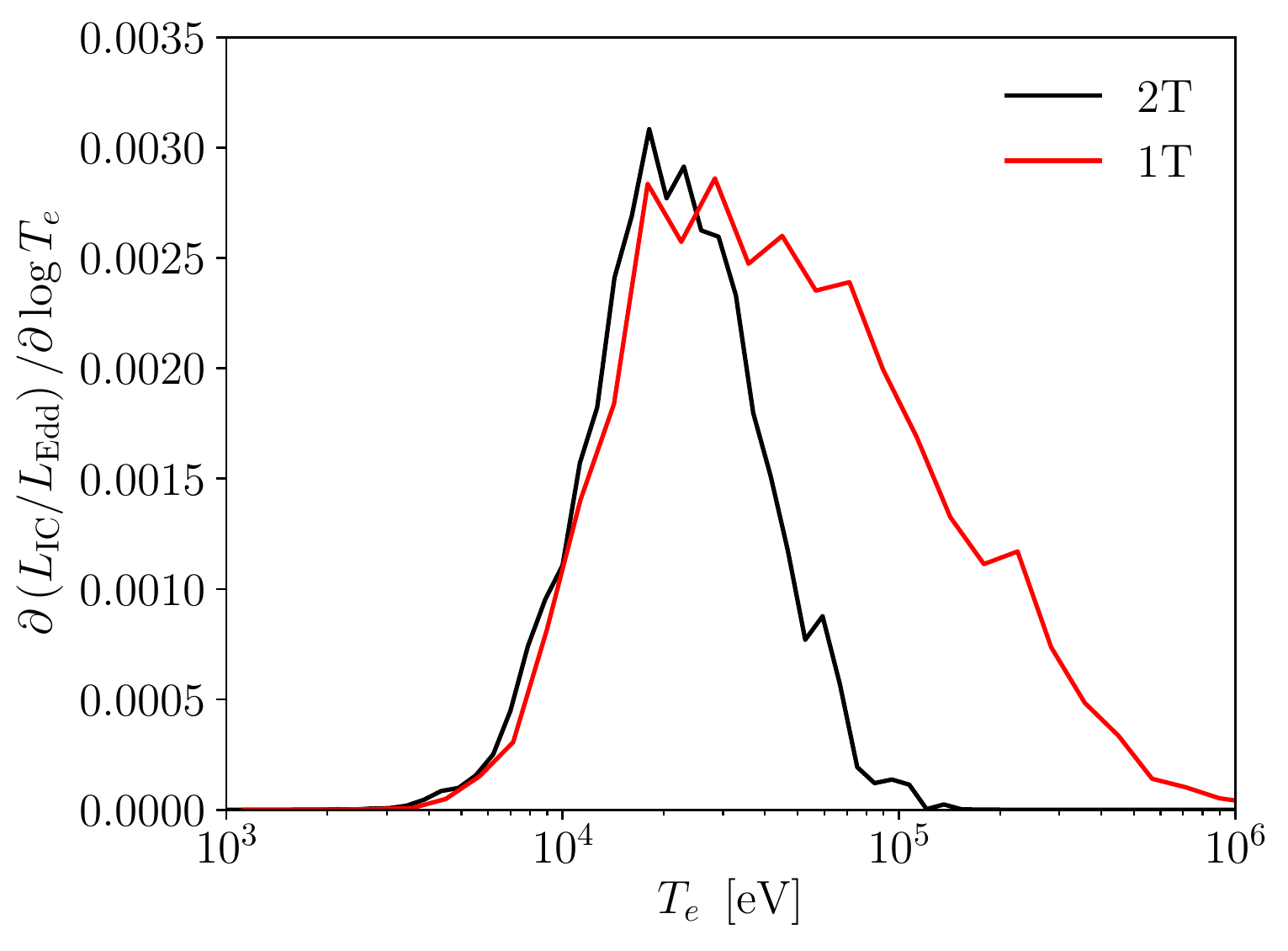}
\caption{The distribution of coronal cooling with respect to the gas electron temperature, for the 2T and 1T simulations each at $t = 1000M$. \label{fig:dLdT_2T}}
\end{figure}

As we did with the 1T simulation, we apply \textsc{pandurata} to successive snapshots of the 2T simulation as well. Figure \ref{fig:dist_obs_spec_2T} shows the spectrum as seen by an observer at infinity, for both the 1T and 2T simulations, each averaged over the range 0--$1000M$. The 2T simulation produces a notably softer X-ray power-law: $\Gamma = 2.53$ compared to the 1T simulation's $\Gamma = 2.25$ (measured on 2--30 keV). In addition, the rollover occurs at a lower energy and falls off more sharply. This is consistent with an overall less-luminous, cooler corona. Compare the distribution of cooling with temperature in Figure \ref{fig:dLdT_2T}: not only is the 2T curve shifted toward cooler temperatures, but the tail above 100 keV in the 1T data---physically located more than $45^\circ$ from the midplane (see Figure \ref{fig:Tes})---is totally absent in the 2T data.

\pagebreak

\section{Conclusion}

We have shown that our simplified calculation of the radiation energy density in the corona required to compute the inverse Compton cooling function is reasonably accurate compared to the much more careful \textsc{pandurata} ray-tracing calculation---especially so in the regions that account for the majority of the coronal cooling. Our confidence in the method is further bolstered by the fact that the time-averaged post-processed continuum spectra are qualitatively similar to real X-ray observations. The power-law index for the 1T simulation run ($\Gamma = 2.25$) and its disk fraction (0.38) place it just barely too soft to meet the standard parameters for a classical ``hard'' spectrum \citep{rem06a}, but it is qualitatively similar. The 2T run results in a power-law that is substantially softer than most X-ray binary spectral observations; to the extent that these simulations adequately capture the effects real coronal physics has on the observed X-rays, we conclude that it is likely that some coupling mechanism stronger than Coulomb collisions is at work in real systems.

We must note, however, that real X-ray spectra of stellar-mass black holes are typically integrated over tens of thousands of seconds; by contrast, the full $2000 M$ run with the new cooling function in place is only 0.01 seconds long. This run cost about 80,000 core-hours to perform on a modern high performance computing cluster. Thus, simulating for a substantial fraction of a real observation is not computationally feasible. It is for exactly this reason, however, that we were motivated to develop a realistic inverse Compton cooling function that does not require the additional computational burden of real transport: \textsc{harm3d} using the IC cooling function in the corona runs at essentially the same speed as with the target-temperature cooling function everywhere, but is considerably more physical. The same cannot be said for \emph{any} treatment involving actual radiation transport.

The success of this method applied to the $a = 0$ case motivates us to apply it to spinning black hole simulations as well. Past efforts to treat the coronal equation of state for black hole accretion have been frustrated by the very large computational expense of true radiation coupling [e.g., \citet{jia19a, jia19b}], while target-temperature
cooling functions without real radiation transfer [e.g., \citet{nob09a, nob10a} and \citet{sha08a, pen10a}] are physically unrealistic. Our new coronal cooling function permits efficient computation while simultaneously providing a very good approximation to the actual cooling rate of coronal plasma; we have also shown that a more realistic coronal cooling function can alter coronal dynamics sufficiently to change the luminosity by tens of percent. Moreover, when these simulations provide the input data for full-up disk-atmosphere plus coronal modeling [via \textsc{pandurata}+\textsc{ptransx}: \citet{kin16a, kin19a}], it becomes possible for the first time to make credible predictions of spectral features, including Fe K$\alpha$ emission, for a wide range of accretion rates onto black holes of all masses and spin parameters.

\acknowledgments

BEK thanks the members of the Center for Theoretical Astrophysics at Los Alamos National Laboratory for helpful conversations, and John G. Baker (NASA GSFC) for useful geometric formulae. BEK was supported by the U.S. Department of Energy Advanced Simulation and Computing Program's Metropolis Fellowship, through the Los Alamos National Laboratory, and used resources provided by the Los Alamos National Laboratory Institutional Computing Program. Los Alamos National Laboratory is operated by Triad National Security, LLC, for the National Nuclear Security Administration of the U.S. Department of Energy (Contract No. 89233218CNA000001). SCN was supported by NSF awards AST-1515982 and OAC-1515969, NASA TCAN award TCAN-80NSSC18K1488, and by an appointment to the NASA Postdoctoral Program at the Goddard Space Flight Center administrated by USRA through a contract with NASA. JDS was supported by NASA TCAN award TCAN-80NSSC18K1488. JHK and BEK were supported by NSF awards AST-1516299, CDI-1028111, and PHYS-1707826.

\bibliography{references}{}

\begin{thebibliography}{}
\expandafter\ifx\csname natexlab\endcsname\relax\def\natexlab#1{#1}\fi
\providecommand{\url}[1]{\href{#1}{#1}}
\providecommand{\dodoi}[1]{doi:~\href{http://doi.org/#1}{\nolinkurl{#1}}}
\providecommand{\doeprint}[1]{\href{http://ascl.net/#1}{\nolinkurl{http://ascl.net/#1}}}
\providecommand{\doarXiv}[1]{\href{https://arxiv.org/abs/#1}{\nolinkurl{https://arxiv.org/abs/#1}}}

\bibitem[{{Abramowicz} {et~al.}(1997){Abramowicz}, {Lanza}, \&
  {Percival}}]{abr97a}
{Abramowicz}, M.~A., {Lanza}, A., \& {Percival}, M.~J. 1997, \apj, 479, 179,
  \dodoi{10.1086/303869}

\bibitem[{{Balbus} \& {Hawley}(1991)}]{bal91a}
{Balbus}, S.~A., \& {Hawley}, J.~F. 1991, \apj, 376, 214,
  \dodoi{10.1086/170270}

\bibitem[{{Balsara} \& {Spicer}(1999)}]{bal99a}
{Balsara}, D.~S., \& {Spicer}, D. 1999, Journal of Computational Physics, 148,
  133, \dodoi{10.1006/jcph.1998.6108}

\bibitem[{{Blumenthal} \& {Gould}(1970)}]{blu70a}
{Blumenthal}, G.~R., \& {Gould}, R.~J. 1970, Reviews of Modern Physics, 42,
  237, \dodoi{10.1103/RevModPhys.42.237}

\bibitem[{{Chandrasekhar}(1960)}]{cha60a}
{Chandrasekhar}, S. 1960, {Radiative transfer}

\bibitem[{{Fragile} {et~al.}(2012){Fragile}, {Gillespie}, {Monahan},
  {Rodriguez}, \& {Anninos}}]{fra12a}
{Fragile}, P.~C., {Gillespie}, A., {Monahan}, T., {Rodriguez}, M., \&
  {Anninos}, P. 2012, \apjs, 201, 9, \dodoi{10.1088/0067-0049/201/2/9}

\bibitem[{{Fragile} {et~al.}(2014){Fragile}, {Olejar}, \& {Anninos}}]{fra14a}
{Fragile}, P.~C., {Olejar}, A., \& {Anninos}, P. 2014, \apj, 796, 22,
  \dodoi{10.1088/0004-637X/796/1/22}

\bibitem[{{Gammie} {et~al.}(2003){Gammie}, {McKinney}, \& {T{\'o}th}}]{gam03a}
{Gammie}, C.~F., {McKinney}, J.~C., \& {T{\'o}th}, G. 2003, \apj, 589, 444,
  \dodoi{10.1086/374594}

\bibitem[{{Haardt} \& {Maraschi}(1991)}]{haa91a}
{Haardt}, F., \& {Maraschi}, L. 1991, \apjl, 380, L51, \dodoi{10.1086/186171}

\bibitem[{{Hawley} \& {Balbus}(1991)}]{haw91a}
{Hawley}, J.~F., \& {Balbus}, S.~A. 1991, \apj, 376, 223,
  \dodoi{10.1086/170271}

\bibitem[{{Jiang} {et~al.}(2019{\natexlab{a}}){Jiang}, {Blaes}, {Stone}, \&
  {Davis}}]{jia19b}
{Jiang}, Y.-F., {Blaes}, O., {Stone}, J.~M., \& {Davis}, S.~W.
  2019{\natexlab{a}}, \apj, 885, 144, \dodoi{10.3847/1538-4357/ab4a00}

\bibitem[{{Jiang} {et~al.}(2014{\natexlab{a}}){Jiang}, {Stone}, \&
  {Davis}}]{jia14a}
{Jiang}, Y.-F., {Stone}, J.~M., \& {Davis}, S.~W. 2014{\natexlab{a}}, \apjs,
  213, 7, \dodoi{10.1088/0067-0049/213/1/7}

\bibitem[{{Jiang} {et~al.}(2014{\natexlab{b}}){Jiang}, {Stone}, \&
  {Davis}}]{jia14b}
---. 2014{\natexlab{b}}, \apj, 796, 106, \dodoi{10.1088/0004-637X/796/2/106}

\bibitem[{{Jiang} {et~al.}(2019{\natexlab{b}}){Jiang}, {Stone}, \&
  {Davis}}]{jia19a}
---. 2019{\natexlab{b}}, \apj, 880, 67, \dodoi{10.3847/1538-4357/ab29ff}

\bibitem[{{Kinch} {et~al.}(2016){Kinch}, {Schnittman}, {Kallman}, \&
  {Krolik}}]{kin16a}
{Kinch}, B.~E., {Schnittman}, J.~D., {Kallman}, T.~R., \& {Krolik}, J.~H. 2016,
  \apj, 826, 52, \dodoi{10.3847/0004-637X/826/1/52}

\bibitem[{{Kinch} {et~al.}(2019){Kinch}, {Schnittman}, {Kallman}, \&
  {Krolik}}]{kin19a}
---. 2019, \apj, 873, 71, \dodoi{10.3847/1538-4357/ab05d5}

\bibitem[{{McKinney} {et~al.}(2014){McKinney}, {Tchekhovskoy}, {Sadowski}, \&
  {Narayan}}]{mck14a}
{McKinney}, J.~C., {Tchekhovskoy}, A., {Sadowski}, A., \& {Narayan}, R. 2014,
  \mnras, 441, 3177, \dodoi{10.1093/mnras/stu762}

\bibitem[{{Mignone} \& {McKinney}(2007)}]{mig07a}
{Mignone}, A., \& {McKinney}, J.~C. 2007, \mnras, 378, 1118,
  \dodoi{10.1111/j.1365-2966.2007.11849.x}

\bibitem[{{Noble} {et~al.}(2009){Noble}, {Krolik}, \& {Hawley}}]{nob09a}
{Noble}, S.~C., {Krolik}, J.~H., \& {Hawley}, J.~F. 2009, \apj, 692, 411,
  \dodoi{10.1088/0004-637X/692/1/411}

\bibitem[{{Noble} {et~al.}(2010){Noble}, {Krolik}, \& {Hawley}}]{nob10a}
---. 2010, \apj, 711, 959, \dodoi{10.1088/0004-637X/711/2/959}

\bibitem[{{Noble} {et~al.}(2011){Noble}, {Krolik}, {Schnittman}, \&
  {Hawley}}]{nob11a}
{Noble}, S.~C., {Krolik}, J.~H., {Schnittman}, J.~D., \& {Hawley}, J.~F. 2011,
  \apj, 743, 115, \dodoi{10.1088/0004-637X/743/2/115}

\bibitem[{{Noble} {et~al.}(2012){Noble}, {Mundim}, {Nakano}, {Krolik},
  {Campanelli}, {Zlochower}, \& {Yunes}}]{nob12a}
{Noble}, S.~C., {Mundim}, B.~C., {Nakano}, H., {et~al.} 2012, \apj, 755, 51,
  \dodoi{10.1088/0004-637X/755/1/51}

\bibitem[{{Novikov} \& {Thorne}(1973)}]{nov73a}
{Novikov}, I.~D., \& {Thorne}, K.~S. 1973, in Black Holes (Les Astres Occlus),
  ed. C.~{Dewitt} \& B.~S. {Dewitt}, 343--450

\bibitem[{{Penna} {et~al.}(2010){Penna}, {McKinney}, {Narayan}, {Tchekhovskoy},
  {Shafee}, \& {McClintock}}]{pen10a}
{Penna}, R.~F., {McKinney}, J.~C., {Narayan}, R., {et~al.} 2010, \mnras, 408,
  752, \dodoi{10.1111/j.1365-2966.2010.17170.x}

\bibitem[{{Porth} {et~al.}(2019){Porth}, {Chatterjee}, {Narayan}, {Gammie},
  {Mizuno}, {Anninos}, {Baker}, {Bugli}, {Chan}, {Davelaar}, {Del Zanna},
  {Etienne}, {Fragile}, {Kelly}, {Liska}, {Markoff}, {McKinney}, {Mishra},
  {Noble}, {Olivares}, {Prather}, {Rezzolla}, {Ryan}, {Stone}, {Tomei},
  {White}, {Younsi}, {Akiyama}, {Alberdi}, {Alef}, {Asada}, {Azulay}, {Baczko},
  {Ball}, {Balokovi{\'c}}, {Barrett}, {Bintley}, {Blackburn}, {Boland},
  {Bouman}, {Bower}, {Bremer}, {Brinkerink}, {Brissenden}, {Britzen},
  {Broderick}, {Broguiere}, {Bronzwaer}, {Byun}, {Carlstrom}, {Chael},
  {Chatterjee}, {Chen}, {Chen}, {Cho}, {Christian}, {Conway}, {Cordes},
  {Geoffrey}, {Crew}, {Cui}, {De Laurentis}, {Deane}, {Dempsey}, {Desvignes},
  {Doeleman}, {Eatough}, {Falcke}, {Fish}, {Fomalont}, {Fraga-Encinas},
  {Freeman}, {Friberg}, {Fromm}, {G{\'o}mez}, {Galison}, {Garc{\'\i}a},
  {Gentaz}, {Georgiev}, {Goddi}, {Gold}, {Gu}, {Gurwell}, {Hada}, {Hecht},
  {Hesper}, {Ho}, {Ho}, {Honma}, {Huang}, {Huang}, {Hughes}, {Ikeda}, {Inoue},
  {Issaoun}, {James}, {Jannuzi}, {Janssen}, {Jeter}, {Jiang}, {Johnson},
  {Jorstad}, {Jung}, {Karami}, {Karuppusamy}, {Kawashima}, {Keating},
  {Kettenis}, {Kim}, {Kim}, {Kim}, {Kino}, {Koay}, {Patrick}, {Koch}, {Koyama},
  {Kramer}, {Kramer}, {Krichbaum}, {Kuo}, {Lauer}, {Lee}, {Li}, {Li},
  {Lindqvist}, {Liu}, {Liuzzo}, {Lo}, {Lobanov}, {Loinard}, {Lonsdale}, {Lu},
  {MacDonald}, {Mao}, {Marrone}, {Marscher}, {Mart{\'\i}-Vidal}, {Matsushita},
  {Matthews}, {Medeiros}, {Menten}, {Mizuno}, {Moran}, {Moriyama},
  {Moscibrodzka}, {M{\"u}ller}, {Nagai}, {Nagar}, {Nakamura}, {Narayanan},
  {Natarajan}, {Neri}, {Ni}, {Noutsos}, {Okino}, {Oyama}, {{\"O}zel},
  {Palumbo}, {Patel}, {Pen}, {Pesce}, {Pi{\'e}tu}, {Plambeck}, {PopStefanija},
  {Preciado-L{\'o}pez}, {Psaltis}, {Pu}, {Ramakrishnan}, {Rao}, {Rawlings},
  {Raymond}, {Ripperda}, {Roelofs}, {Rogers}, {Ros}, {Rose}, {Roshanineshat},
  {Rottmann}, {Roy}, {Ruszczyk}, {Rygl}, {S{\'a}nchez},
  {S{\'a}nchez-Arguelles}, {Sasada}, {Savolainen}, {Schloerb}, {Schuster},
  {Shao}, {Shen}, {Small}, {Sohn}, {SooHoo}, {Tazaki}, {Tiede}, {Tilanus},
  {Titus}, {Toma}, {Torne}, {Trent}, {Trippe}, {Tsuda}, {van Bemmel}, {van
  Langevelde}, {van Rossum}, {Wagner}, {Wardle}, {Weintroub}, {Wex}, {Wharton},
  {Wielgus}, {Wong}, {Wu}, {Young}, {Young}, {Yuan}, {Yuan}, {Zensus}, {Zhao},
  {Zhao}, {Zhu}, \& {Event Horizon Telescope Collaboration}}]{eht19a}
{Porth}, O., {Chatterjee}, K., {Narayan}, R., {et~al.} 2019, \apjs, 243, 26,
  \dodoi{10.3847/1538-4365/ab29fd}

\bibitem[{{Remillard} \& {McClintock}(2006)}]{rem06a}
{Remillard}, R.~A., \& {McClintock}, J.~E. 2006, \araa, 44, 49,
  \dodoi{10.1146/annurev.astro.44.051905.092532}

\bibitem[{{Roedig} {et~al.}(2012){Roedig}, {Zanotti}, \& {Alic}}]{roe12a}
{Roedig}, C., {Zanotti}, O., \& {Alic}, D. 2012, \mnras, 426, 1613,
  \dodoi{10.1111/j.1365-2966.2012.21821.x}

\bibitem[{{Ryan} \& {Dolence}(2019)}]{rya19a}
{Ryan}, B.~R., \& {Dolence}, J.~C. 2019, arXiv e-prints, arXiv:1907.09625.
\newblock \doarXiv{1907.09625}

\bibitem[{{Ryan} {et~al.}(2015){Ryan}, {Dolence}, \& {Gammie}}]{rya15a}
{Ryan}, B.~R., {Dolence}, J.~C., \& {Gammie}, C.~F. 2015, \apj, 807, 31,
  \dodoi{10.1088/0004-637X/807/1/31}

\bibitem[{{Rybicki} \& {Lightman}(1986)}]{ryb86a}
{Rybicki}, G.~B., \& {Lightman}, A.~P. 1986, {Radiative Processes in
  Astrophysics}, 400

\bibitem[{{S{\c a}dowski} {et~al.}(2014){S{\c a}dowski}, {Narayan}, {McKinney},
  \& {Tchekhovskoy}}]{sad14a}
{S{\c a}dowski}, A., {Narayan}, R., {McKinney}, J.~C., \& {Tchekhovskoy}, A.
  2014, \mnras, 439, 503, \dodoi{10.1093/mnras/stt2479}

\bibitem[{{S{{\c a}}dowski} {et~al.}(2016){S{{\c a}}dowski}, {Tejeda},
  {Gafton}, {Rosswog}, \& {Abarca}}]{sad16a}
{S{{\c a}}dowski}, A., {Tejeda}, E., {Gafton}, E., {Rosswog}, S., \& {Abarca},
  D. 2016, \mnras, 458, 4250, \dodoi{10.1093/mnras/stw589}

\bibitem[{{Schnittman} \& {Krolik}(2013)}]{sch13b}
{Schnittman}, J.~D., \& {Krolik}, J.~H. 2013, \apj, 777, 11,
  \dodoi{10.1088/0004-637X/777/1/11}

\bibitem[{{Schnittman} {et~al.}(2013){Schnittman}, {Krolik}, \&
  {Noble}}]{sch13a}
{Schnittman}, J.~D., {Krolik}, J.~H., \& {Noble}, S.~C. 2013, \apj, 769, 156,
  \dodoi{10.1088/0004-637X/769/2/156}

\bibitem[{{Shafee} {et~al.}(2008){Shafee}, {McKinney}, {Narayan},
  {Tchekhovskoy}, {Gammie}, \& {McClintock}}]{sha08a}
{Shafee}, R., {McKinney}, J.~C., {Narayan}, R., {et~al.} 2008, \apjl, 687, L25,
  \dodoi{10.1086/593148}

\bibitem[{{Shakura} \& {Sunyaev}(1973)}]{sha73a}
{Shakura}, N.~I., \& {Sunyaev}, R.~A. 1973, \aap, 500, 33

\bibitem[{{Stepney}(1983)}]{ste83a}
{Stepney}, S. 1983, \mnras, 202, 467, \dodoi{10.1093/mnras/202.2.467}

\bibitem[{{Stepney} \& {Guilbert}(1983)}]{ste83b}
{Stepney}, S., \& {Guilbert}, P.~W. 1983, \mnras, 204, 1269,
  \dodoi{10.1093/mnras/204.4.1269}

\bibitem[{{Zanotti} {et~al.}(2011){Zanotti}, {Roedig}, {Rezzolla}, \& {Del
  Zanna}}]{zan11a}
{Zanotti}, O., {Roedig}, C., {Rezzolla}, L., \& {Del Zanna}, L. 2011, \mnras,
  417, 2899, \dodoi{10.1111/j.1365-2966.2011.19451.x}

\end{thebibliography}
\bibliographystyle{aasjournal}

\end{document}